\documentclass[onecolumn,showpacs,amsmath,amssymb,prd,nofootinbib]{revtex4}
\usepackage[usenames]{color}
\usepackage{graphicx}
\usepackage{amsmath}
\usepackage{amssymb}
\usepackage{amsfonts}

\newcommand{\bc}{\begin{center}}
\newcommand{\ec}{\end{center}}
\newcommand{\bt}{\begin{tabular}}
\newcommand{\et}{\end{tabular}}
\newcommand{\be}{\begin{equation}}
\newcommand{\ee}{\end{equation}}
\newcommand{\bea}{\begin{eqnarray}}
\newcommand{\eea}{\end{eqnarray}}
\newcommand{\bfig}{\begin{figure}}
\newcommand{\efig}{\end{figure}}
\newcommand{\nn}{\nonumber}
\newcommand{\raw}{\rightarrow}

\newcommand{\YS}{Y^{(S)}}
\newcommand{\YSv}{Y^{(S)}_i}
\newcommand{\YSt}{Y^{(S)}_{ij}}

\newcommand{\gt}{\gamma_{ij}}

\newcommand{\Hu}{{\mathcal{H}}}

\def\gsim{ \lower .75ex \hbox{$\sim$} \llap{\raise .27ex \hbox{$>$}} }
\def\lsim{ \lower .75ex \hbox{$\sim$} \llap{\raise .27ex \hbox{$<$}} }

\begin{document}

\title {Biases on cosmological parameters by general relativity effects}

\author{L.~Lopez-Honorez$^1$, O.~Mena$^2$ and S.~Rigolin$^3$}

\begin{abstract}
General relativistic corrections to the  galaxy power spectrum appearing
at the horizon scale, if neglected, may induce biases on the measured values 
of the cosmological parameters. In this paper, we study the impact of general 
relativistic effects on non standard cosmologies such as scenarios with a time 
dependent dark energy equation of state, with a coupling between the dark energy 
and the dark matter fluids or with non--Gaussianities.  We then explore whether 
general relativistic corrections affect future constraints on cosmological parameters 
in the case of a constant dark energy equation of state and of non--Gaussianities. 
We find that relativistic corrections on the power spectrum are not expected to affect 
the foreseen errors on the cosmological parameters nor to induce large biases on them. 
\end{abstract}
\affiliation{$^1$ Service de Physique Th\'eorique, Universit\'e Libre de Bruxelles, Brussels, Belgium}
\affiliation{$^2$ IFIC-CSIC and Universidad de Valencia, Valencia, Spain}
\affiliation{$^3$ Dipartimento di Fisica, Universit\`a di Padova and  
                       INFN Padova, Padova, Italy}

\begin{flushright}
DFPD-11/TH/13\\
{ULB-TH/11-24}\\
	IFIC/11-51

\end{flushright}

\maketitle

\setcounter{footnote}{0}

\section{Introduction}
\label{sec:introduction}

The complete general relativistic description of the observed matter power spectrum is, 
at large scales, significantly different from the standard Newtonian one. The observed 
redshift and position of galaxies are affected by matter fluctuations and gravity 
waves between the source and the observer, see e.g. Ref.~\cite{Bonvin:2005ps}. In addition, 
the matter density perturbation, $\delta_{\rm m}$, is gauge dependent while observable quantities, 
such as the power spectrum, should be gauge invariant. The standard picture looks, therefore, 
incomplete, and a general relativistic description is needed in order to correctly compute the 
measured observables~\cite{Yoo:2009au,Yoo:2010ni}. Current observations, based on available 
galaxy surveys, are not affected, in practice, by general relativistic corrections since 
they appear only at very large scales. In future galaxy surveys, however, these corrections 
may interfere with the measure of other physical effects which modify the 
large-scale shape of the power spectrum. 

In this paper,  we study the  general relativistic effects in several 
cosmological scenarios, like: i) a constant dark energy equation of state; ii) a time varying 
equation of state $w(a)$; iii) non--Gaussianities; iv) a coupling
between dark energy and dark matter;  and finally iv) massive 
neutrinos. For the scenarios i) and
iii),  we compute the expected 
errors and biases from a future
Euclid-like galaxy survey by means of a Fisher matrix analysis,
comparing the results with and without general relativistic corrections in the matter power spectrum. 

The structure of the paper is the following. Section \ref{sec:seci} summarizes the general 
relativistic corrections treatment. In Sec.~\ref{sec:models} the impact of general relativistic 
corrections in the cosmological scenarios quoted above is
presented. The expected errors and  biases on the cosmological
parameters are computed  in Sec.~\ref{sec:fisher} for two particular scenarios. Finally in Sec.~\ref{sec:concl} we conclude.  

\section{Preliminaries}
\label{sec:seci}
Following the results of Refs.~\cite{Bonvin:2005ps,Yoo:2009au,Yoo:2010ni,Bonvin:2011bg,Challinor:2011bk}, 
we briefly summarize the treatment of the observed galaxy power spectrum in redshift space. In linear 
perturbation theory, the observed (matter) density $\rho_{\rm m}$ at a given redshift is defined 
as a function of the density fluctuation $\delta_{\rm m}$ and  the background (matter) density 
$\bar{\rho}_{\rm m}$
\begin{eqnarray}
\rho_{\rm m}  &\equiv& \bar{\rho}_{\rm m} (\bar{z}) (1+\delta_{\rm m}) \, .
\label{eq:1}
\end{eqnarray}
For the standard $\Lambda$CDM cosmology, that we take as reference, the background matter density 
in terms of the current Hubble parameter $H_0$ and today's matter density relative to the critical 
density $\Omega_{\rm m}$ reads~\footnote{See Sec.~\ref{sec:coupl-modif-grav} for non standard 
cosmologies in which Eq.~(\ref{eq:2}) is not valid.}:
\begin{equation}
\bar{\rho}_{\rm m}(\bar{z})=\frac{3 H_0^2}{8 \pi G}\Omega_{\rm m}(1+\bar{z})^3 \, .
\label{eq:2}
\end{equation}
In Eqs.~(\ref{eq:1}) and (\ref{eq:2}), the dependence of the background density $\bar{\rho}_{\rm m}$ 
on the background redshift $\bar z$ has been made explicit. The ratio between the emitter and the 
observed frequencies at the background level is defined as
\begin{eqnarray}
1 + \bar z = \frac{\bar \nu_e}{\bar \nu_o} = 
\frac{\left(\bar K^\mu \bar u_\mu\right)_{e}}{\left(\bar K^\mu \bar u_\mu\right)_{o}} = \frac{1}{\bar a} \, ,
\end{eqnarray}
where $\bar K^\mu$ and $\bar u_{\mu}$ are the background photon wave vector and the background 
emitter/observer ($e/o$) four-velocity, respectively. At linear order in perturbation theory, 
the observed redshift of a given source, $z$, differs from the background one due to the
 matter/gravity fluctuations that the photon encounters between the emitter and the observer 
positions. The perturbed four-velocity and photon null--vector read:
\begin{eqnarray}
u^\mu &=& \frac{1}{\bar a} \left( 1 - A ,  v^i \right) \, ; \\ 
K^\mu &=& \frac{\bar \nu}{\bar a} \left( 1 + \frac{\delta \nu}{\bar \nu} ,  n^i + \delta n^i\right) \,,
\label{eq:pertnull}
\end{eqnarray}
where $\mu=0,..,3$ and $i=1,..,3$. $v^i$ is the peculiar velocity of the observer/emitter and 
$\delta \nu$ and $\delta n^i$ are the perturbed photon frequency and propagation direction, respectively. 
The conventions used for the perturbed Friedmann-Robertson-Walker (FRW) metric, together with a list of 
useful relations, can be found in Appendix~\ref{sec:gauge-invar-form}. The observed (perturbed) redshift  
$z$ thus reads:
\begin{eqnarray}
1 + z \equiv 1 + \bar z + \delta z = 
\frac{\left(K^\mu u_\mu\right)_{e}}{\left(K^\mu u_\mu\right)_{o}} = \left(1 + \bar z\right) 
\left[ 1+ \left(\frac{\delta \nu}{\bar \nu} + A + \left(B_i-v_i \right) n^i\right)^{e}_{o} \right] \, .
\label{eq:obsz}
\end{eqnarray}

Expressing the matter density in terms of the observed redshift, instead of the unobservable 
background one, it gives:
\begin{eqnarray}
\rho_{\rm m}  = \bar{\rho}_{\rm m} (z)
\left(1+\delta_{\rm m} - \frac{ d\bar{\rho}_{\rm m}}{dz} \frac{\delta z}{\bar{\rho}_{\rm m} }\right)  
 \equiv \bar{\rho}_{\rm m} (z) \left(1+\Delta_z\right) \, ,
\label{eq:rhom}
\end{eqnarray}
with the background matter density $\bar\rho_m(z)$ function of the observed redshift $z$. While the density 
contrast $\delta_{\rm m}$ and the redshift fluctuation $\delta z$ are gauge dependent quantities, their 
combination $\Delta_z$ is, instead, gauge invariant. Notice, however, that the truly observed quantity 
is the galaxy number density perturbation~\cite{Yoo:2009au,Bonvin:2011bg,Challinor:2011bk}
corresponding to:
\begin{equation}
\Delta_{obs}=\frac{\delta N}{\overline N} \equiv \frac{N(z)-\overline N(z)}{\overline N(z)} = \Delta_z + 
\frac{\delta {\rm Vol}}{\overline{{\rm Vol}}}~,
\end{equation}
where an extra contribution from the physical survey volume perturbation appears. Being the volume  
density perturbation, $\delta {\rm Vol}$, a gauge invariant quantity, $\Delta_{obs}$ is automatically 
gauge invariant, as it should be for any observable quantity. In addition, one has to introduce a bias 
between galaxy and matter overdensities.  We will ignore for the moment the bias issue, deferring a brief 
discussion of this aspect to Sec.~\ref{sec:non-gaussianity}.

Making use of the null energy condition and the photon geodesic equation (see 
Appendix~\ref{sec:gauge-invar-form} and also 
Refs.~\cite{Bonvin:2005ps,Yoo:2009au,Yoo:2010ni,Bonvin:2011bg,Jeong:2011as} for more details) 
one can write $\Delta_z$ in terms of gauge invariant quantities as:
\begin{equation}
\Delta_z= \Delta_{\rm m}+ 3\, \mathbf{ n \cdot V }+3 \left(\Psi_B-\Phi_B\right)-3\int_{\lambda_o}^{\lambda_s} d\lambda
\left(\dot \Psi_B-\dot \Phi_B\right)\,, 
\label{eq:Deltaz}
\end{equation}
where $\Phi_B$ and $\Psi_B$ are the Bardeen potentials and $\Delta_{\rm m}$ and $\mathbf V$ are the 
gauge invariant matter density contrast  and   peculiar velocity, which definitions can be found in 
Appendix~\ref{sec:gauge-invar-form}. The last term in Eq.~(\ref{eq:Deltaz}) is the usual integrated 
Sachs--Wolfe effect between the observer and the emission point with $d/d\lambda=\partial_{\tau}+n^i\partial_i$. 
The survey volume perturbation $\delta {\rm Vol}$ has been carefully derived in several references, see
e.g. Refs.~\cite{Yoo:2009au,Bonvin:2011bg}, we therefore omit the details of its calculation here. 
Neglecting the unmeasurable monopole and dipole perturbations at the observer position, the expression
of $\Delta_{obs}$ in terms of gauge invariant quantities  reads: 
\begin{eqnarray}
\Delta_{obs} &=&\Delta_{\rm m}+\Psi_B-\Phi_B- \mathbf{n\cdot V} -\frac{1}{\cal H}\left[n^i\partial_i\Psi_B+\dot 
 \Phi_B +\frac{d}{d\lambda} (\mathbf{n\cdot V}) \right]\cr
 &&+\left(\frac{2}{r_s{\cal H}}+\frac{\dot {\cal H}}{{\cal H}^2}\right)\left[ \mathbf{n\cdot V}+
  \Psi_B+\int_0^{r_s} d\lambda \left(\dot\Psi_B-\dot\Phi_B\right)\right]\cr 
 &&+\frac{2}{r_s}\int_0^{r_s} d\lambda
  \left(\Psi_B-\Phi_B\right)-\frac{1}{r_s}\int_0^{r_s} d\lambda
  \frac{r_s-r}{r} \Delta_\Omega\left(\Psi_B-\Phi_B\right)~,
  \label{eq:DeltaDur}
\end{eqnarray}
where $r_s=\int_{\tau_o}^{\tau_s} d\tau$ corresponds to the comoving distance between the source and 
the observer and $\Delta_\Omega$ is the angular Laplacian on a unit
sphere.
Notice that Eq.~(\ref{eq:DeltaDur}) holds for the standard cosmology case and reduces to Eq.~(30) of
Ref.~\cite{Bonvin:2011bg} once the Euler equation for the gauge invariant matter velocity scalar 
perturbation:
\begin{equation}
  \dot V^i= -{\cal H} V^i-\partial^i \Psi_B\,,
\label{eq:Eulerst}
\end{equation}
is implemented. Also, let us emphasize that we have assumed a constant
comoving source number density  and ignored the vector and tensor
contributions in Eqs.~(\ref{eq:Deltaz}) and~(\ref{eq:DeltaDur}).

For later convenience, let us express Eqs.~(\ref{eq:Deltaz}) and (\ref{eq:DeltaDur}) in the Newtonian 
gauge. The density perturbations $\Delta_z$ and $\Delta_{obs}$ read respectively:
\begin{eqnarray}
\Delta_z &=&\delta_{\rm m}^N + 3\, \mathbf{n\cdot v}+3\Psi_N-3\int d\lambda(\dot
  \Psi_N+\dot\Phi_N)\label{eq:DzN}\,;\\
   \cr
\Delta_{obs} &=&\delta_{\rm m}^N+\frac{1}{\cal H} \mathbf{n \cdot} \partial_r \mathbf{v}-2\kappa 
 +\Psi_N-2\Phi_N +\frac{1}{\cal H} \dot \Phi_N\cr
&&+\left(\frac{2}{r_s{\cal H}}+\frac{\dot {\cal H}}{{\cal H}^2}\right)\left[ \mathbf{n \cdot v}+\Psi_N+
  \int_0^{r_s} d\lambda \left(\dot\Psi_N+\dot\Phi_N\right)\right]+
  \frac{2}{r_s}\int_0^{r_s} d\lambda\left(\Psi_N+\Phi_N\right)~,
\label{eq:DobsN}
\end{eqnarray}
where $\kappa$ is the lensing convergence (see Eq.~(\ref{eq:kappa})), $\Psi_N$ and $\Phi_N$ are the scalar 
perturbations of the metric in the Newtonian gauge (see Appendix~\ref{sec:gauge-invar-form}) and the partial  
derivative $\partial_r = e_r^i\partial_i=-n^i\partial_i$ with $e_r^i$ indicating the source position. 
With $\delta_m^N$  and $\mathbf v$, we refer to the matter density and peculiar velocity perturbation in the
Newtonian gauge. 

In the standard Newtonian approximation, the galaxy number density
perturbation, $\Delta_{st}$, only gets contributions from the three
first terms of Eq.~(\ref{eq:DobsN}), namely from  the matter density
perturbation, the redshift space distortion term and from the
convergence term. We consider that, neglecting the bias between galaxy
and matter overdensities,  the associated standard
Newtonian power  
spectrum is  related to the matter power spectrum evaluated in the
synchronous gauge\footnote{For a comprehensive discussion see for
  example Refs.~\cite{Challinor:2011bk,Jeong:2011as}.} in the following way: 
\begin{equation}
P_{\Delta_{st}}=P^S_{\rm m}\left(1+f_{\rm eff}\mu_k^2\right)^2~.
\label{eq:Pst}
\end{equation}
The latter is typically   
used for calculating the power spectrum when relativistic
contributions can be safely neglected (i.e. for scales much smaller
than the horizon scale). In Eq.~(\ref{eq:Pst}), the index $S$ refers
to the synchronous comoving gauge,
 $f_{\rm eff}$ is the linear growth function and $\mu_k$ is the 
cosine of the angle   
between the line of sight and the wave vector $k$. In standard
cosmological scenarios, the growth function  
$f_{\rm eff}$ is given by $d\ln\delta_{\rm m}/d\ln a$. 

Notice that in Eq.~(\ref{eq:Pst}) we have ignored the contribution
from the convergence term.  
Through all this study contributions from projected quantities have been neglected when computing the 
3-D power spectrum $P(k)$, while their contributions have been accounted for when calculating the 2-D 
angular power spectrum $C_\ell$. Also, given that we expect galaxy formation to proceed in the potential 
wells of dark matter halos\footnote{We know that luminous red galaxies occupy massive dark matter halos 
today from weak lensing measurements~\cite{Mandelbaum:2005nx}.}, all the computed power spectra 
($P(k)$ or $C_\ell$) in the following will correspond to the dark matter power spectra  
which have been obtained using the available public version of {\tt CAMB}~\cite{camb}.

\section{Cosmological scenarios}
\label{sec:models}
We explore below the impact of general relativistic corrections in
several cosmological scenarios which include the presence of a constant 
dark energy equation of state and the presence of non--Gaussianities. In
the next section, we will estimate the foreseen errors on the several
cosmological parameters involved in each of these two cosmologies
 using the Fisher matrix formalism. For the sake of illustration,
 we also discuss the effect of  general relativistic corrections on 
the observed galaxy power spectrum in the case of a time varying  
equation of state $w(a)$, a coupling
between dark energy and dark matter and  massive 
neutrinos. 

Unless otherwise stated, the following numerical values for the cosmological parameters have been used: 
$\Omega_{\rm b} h^2=0.02267$, $\Omega_{\rm dm} h^2=0.1131$, $h=0.705$, the scalar amplitude $A_{\rm s}=2.64 
\times 10^{-9}$ and the scalar spectral index $n_{\rm s}=0.96$. $\Omega_{\rm{b(dm)}}$ refers 
to the current baryon (dark matter) energy density relative to the critical density and $h$ is related to 
the present value of the Hubble parameter $H_0=100 h$~Mpc/km/s. The sound speed for the dark energy fluid 
is fixed to $c_{\rm s}^2=1$.  

\subsection{Dark energy}
\label{sec:dark-energy}

\begin{figure}[tb!]
\vspace{-0.1cm}
\begin{center}
\begin{tabular}{cc}
\includegraphics[width=8cm]{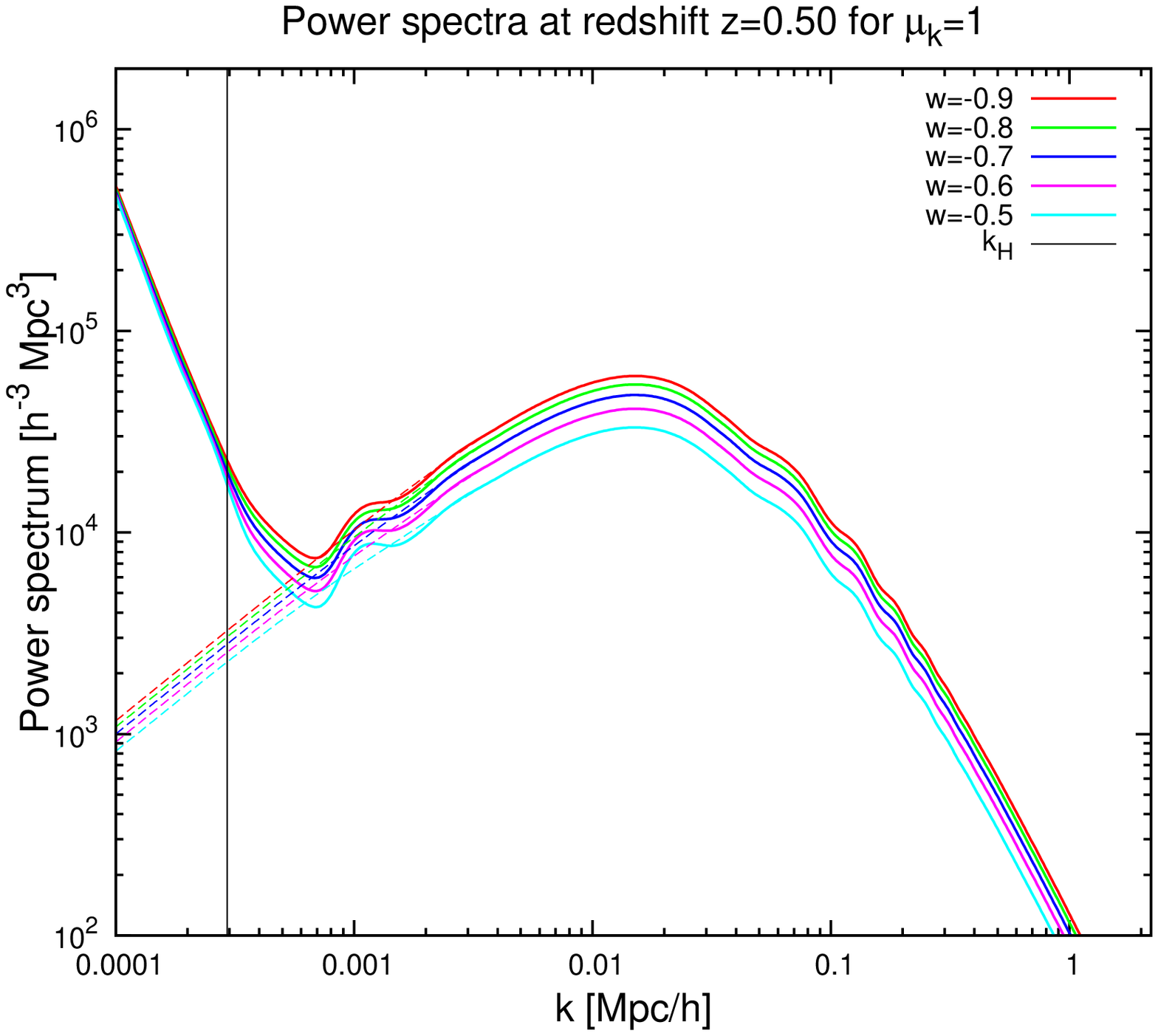}&
\includegraphics[width=8cm]{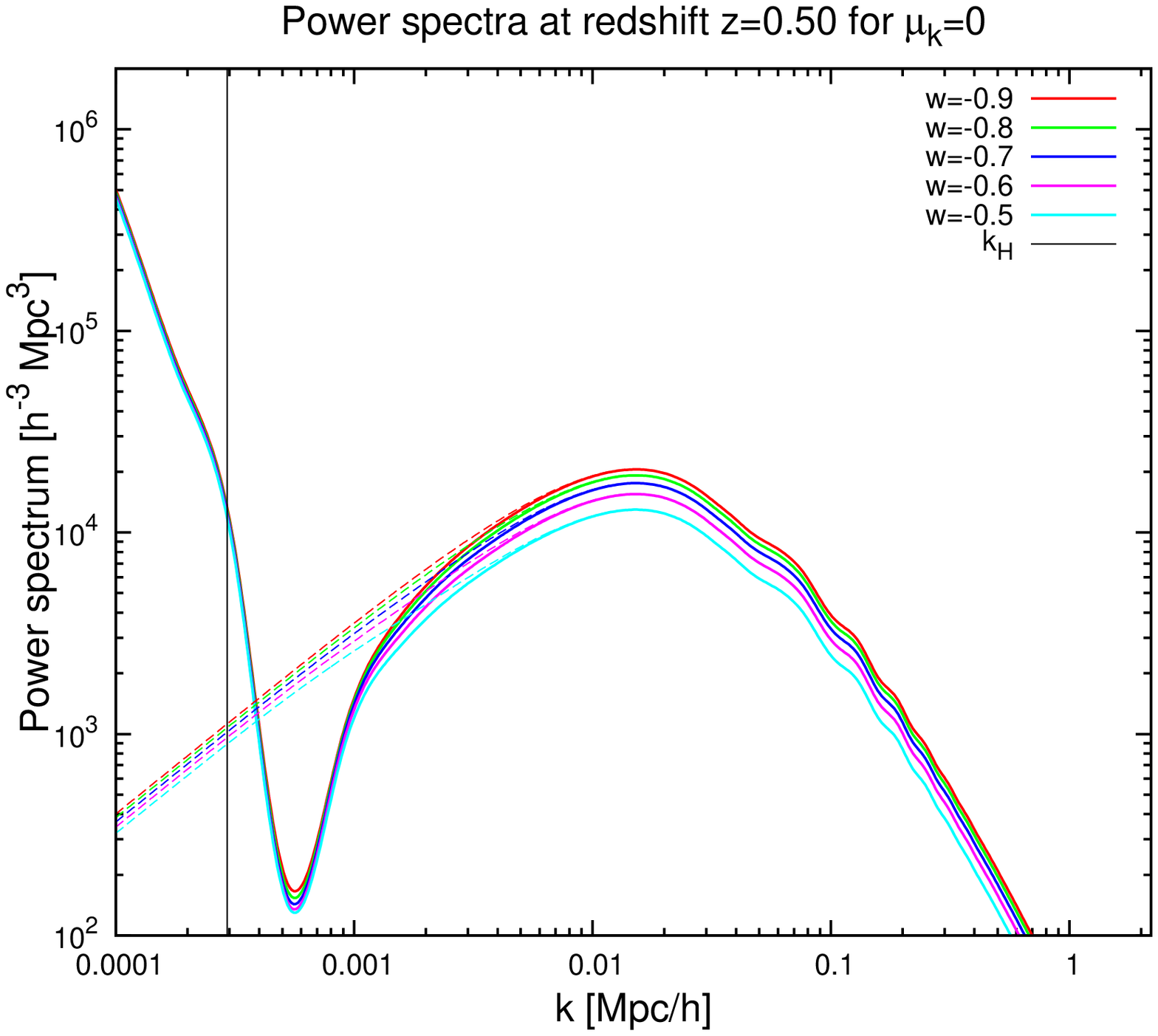}
\end{tabular}
\caption{\it 
  $P_{\Delta_{obs}}(k)$ (solid lines) and  $P_{\Delta_{st}}(k)$ (dashed lines) for $\mu_k=1$ (left panel) 
  and $\mu_k=0$ (right panel). The dark energy equation of state $w$ is assumed to be constant and has been varied between 
  $-0.9$ and $-0.5$. The vertical line corresponds to the horizon scale $k_H$ for $w=-0.9$.}
\label{fig:Pw}
\end{center}
\end{figure}

We first consider a cosmological model including standard cold dark matter and a dark energy 
fluid characterized by a constant equation of state $w$. Figure \ref{fig:Pw} shows the dark 
matter power spectra $P_{\Delta_{obs}}(k,\mu_k)$ and $P_{ \Delta_{st} }(k,\mu_k)$ for both 
the line-of-sight ($\mu_k=1$) and the transverse ($\mu_k=0$) modes at $z=0.5$ for several 
values of $w$, ranging from $w=-0.9$ to $w=-0.5$. The horizon scale, $k_H$ is also shown 
for the $w=-0.9$ case. 
Notice that, in  Sec.~\ref{sec:neutrino-masses}, we discuss
the $k$ position of the dip appearing at large scales in the power spectrum $P_{\Delta_{obs}}(k,\mu_k)$
in the transverse direction (right plot).
The modifications in the shape of the
power spectra when relativistic effects are considered barely change
when the dark energy equation of state is varied. In addition, the new
features on the power spectrum induced by the general relativity terms
appear only at very large scales: consequently, one would not expect much
improvement on the measurement of $w$ when relativistic effects are
included. For the same reason, the bias induced on the dark energy
equation of state $w$ when the data are fitted to $P_{\Delta_{st} }$ 
(instead of using the full description given by $P_{\Delta_{obs}}$) is expected to 
be negligible. In Sec.~\ref{sec:fisher}, we estimate the foreseen errors on several
cosmological parameters within a constant dark energy equation of state cosmological 
scenario using the Fisher matrix formalism.

We also consider a time varying equation of state with a parameterization that has been extensively explored 
in the literature~\cite{Chevallier:2000qy,Linder:2002et,Albrecht:2006um,Linder:2006sv}: $w(a)=w_0+w_a(1-a)$. 
We study 3 cases: i) $w_0=-1$ and $w_a=0$; ii) $w_0=-1$ and $w_a=1$;
iii) $w_0=-1$ and $w_a=-1$. For our numerical calculations, we have used the Parameterized Post-Friedmann (PPF) prescription for the dark energy perturbations, see Refs. \cite{Hu:2007pj,Hu:2008zd,Fang:2008sn}.

\begin{figure}[tb!]
\vspace{-0.1cm}
\begin{center}
\begin{tabular}{cc}
\includegraphics[width=8cm]{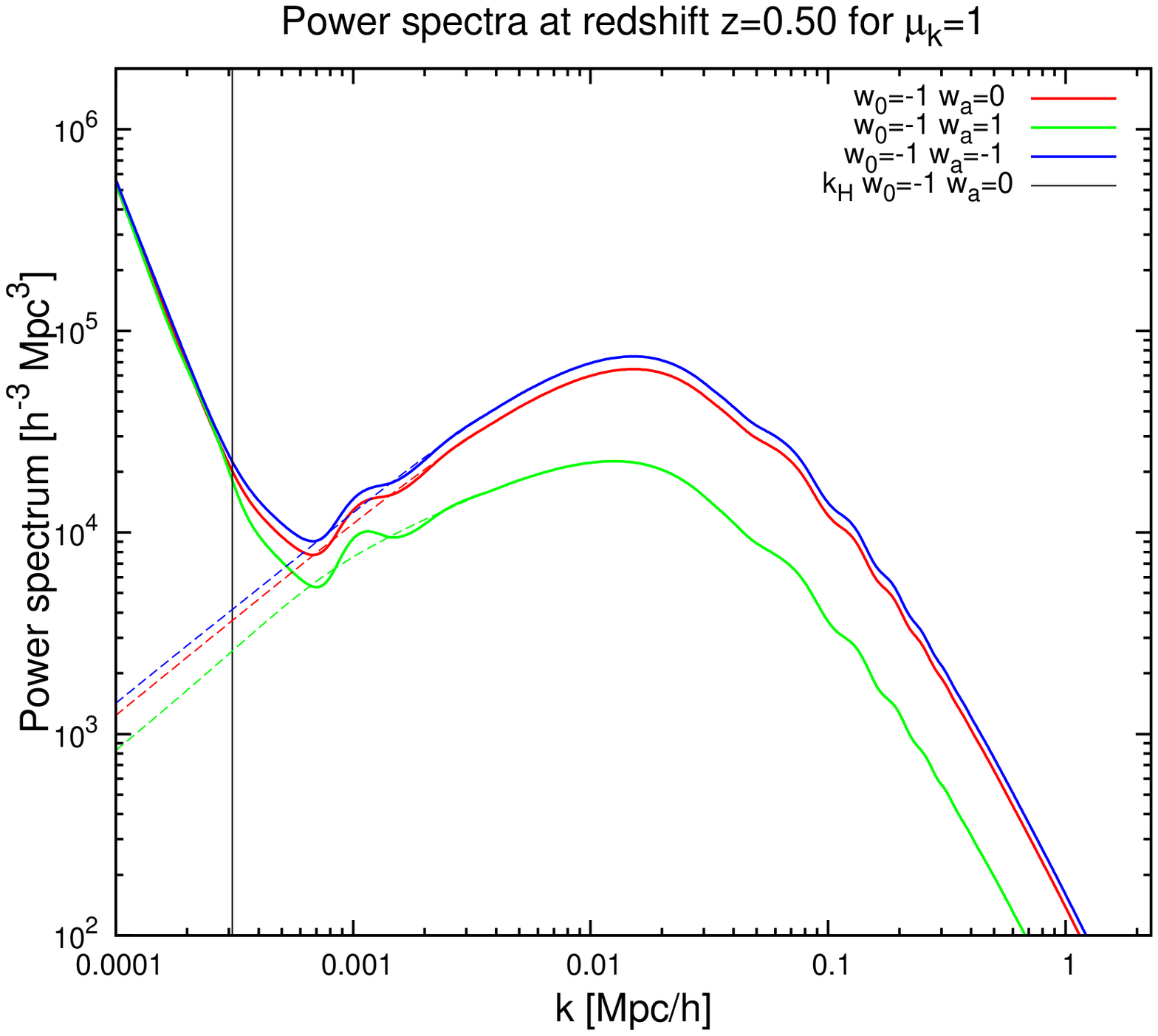}&
\includegraphics[width=8cm]{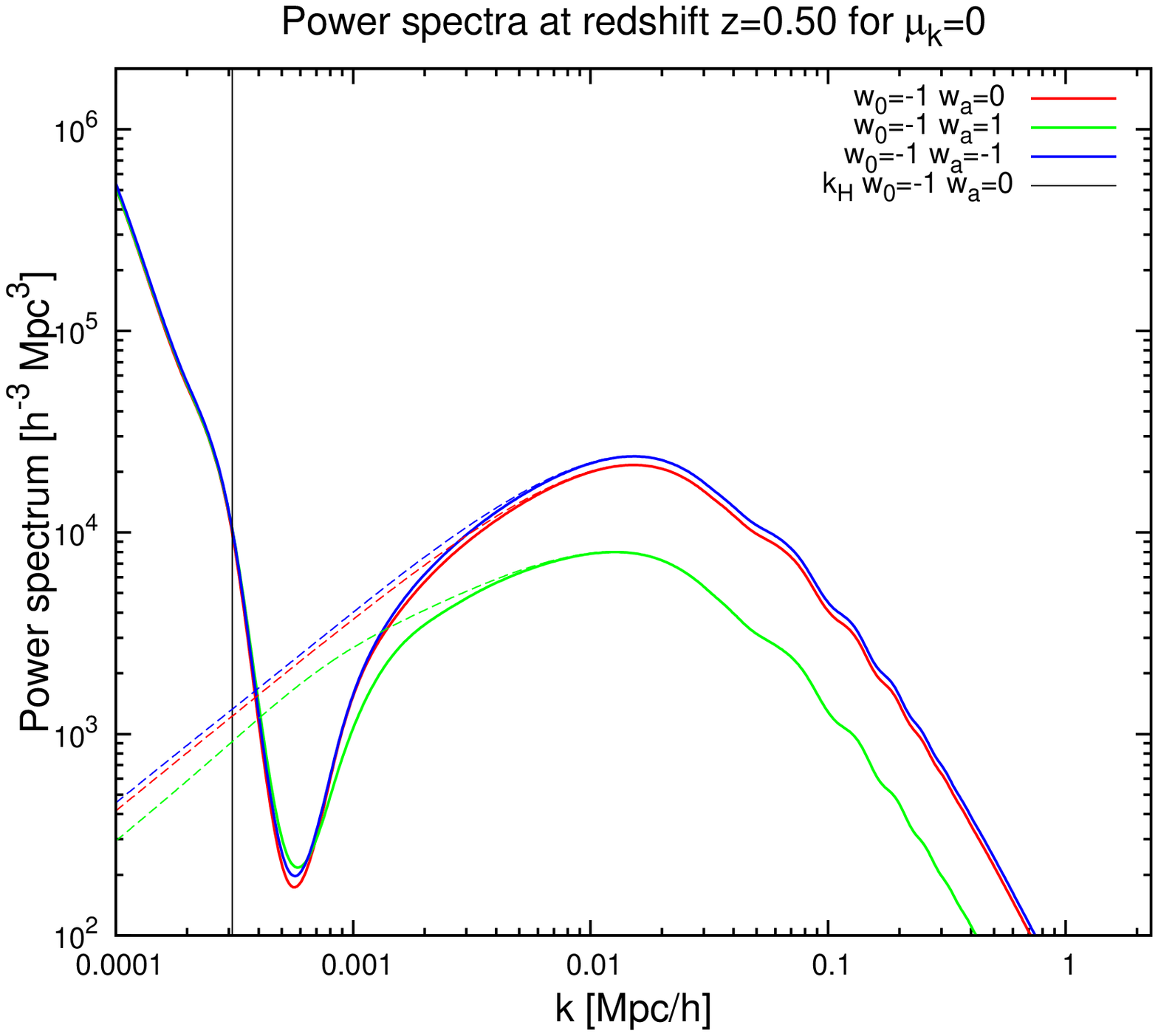}
\end{tabular}
\caption{\it \it $P_{\Delta_{obs}}(k)$ (solid lines) and  $P_{\Delta_{st}}(k)$
  (dashed lines) for $\mu_k=1$ (left panel) and $\mu_k=0$ (right panel) for
  the three possible $w(a)$ cosmologies explored here at $z=0.5$.
 The vertical lines depict the horizon scale $k_H$ for $w_0=-1$ and $w_a=0$.}
\label{fig:Pwa}
\end{center}
\end{figure}

Figure \ref{fig:Pwa} shows the standard Newtonian matter power spectrum and the one with relativistic 
corrections included for the three $w(a)$ cosmologies above, for both the line of sight and the transverse 
modes at redshift $z=0.5$. Notice that the dependence of the general relativistic corrections on the time 
varying equation of state $w(a)$ is as mild as for the case of constant $w$. 
The new features due to general relativistic corrections in the line of sight 
modes barely change when different values of $w_0$ and $w_a$ are 
considered. Therefore, the extra information contained in these 
general relativistic terms will poorly increase the precision on 
the measurement of a time varying dark energy equation of state. 

\subsection{Non--Gaussianity}
\label{sec:non-gaussianity}

In this section, we take into account a non zero bias between galaxy and dark matter overdensities. 
Following the prescription of several recent studies~\cite{Challinor:2011bk,Bruni:2011ta,Jeong:2011as} 
and considering a linear bias relation in the comoving synchronous gauge, the galaxy and dark matter 
overdensities are related by $\delta^S_g = b\, \delta^S_{\rm dm}$. If the primordial fluctuations are Gaussian, 
it is generally assumed that this bias $b$ is scale independent. 

Deviations from Gaussian initial conditions offer a unique tool for testing the mechanism which generated 
primordial perturbations. Non--Gaussianities are commonly characterized by a single parameter, $f_{\rm NL}$. 
The local primordial Bardeen gauge-invariant potential on large scales in the matter dominated era can 
be written as~\cite{Salopek:1990jq,Gangui:1993tt,Verde:1999ij,Komatsu:2001rj} 
\begin{equation}
\Phi_{\rm NG}=\Phi_G+f_{\rm NL}\left(\Phi_{\rm G}^2-\langle \Phi_{\rm G}^2 \rangle\right)\,,
\label{eq:fnl}
\end{equation}
where $\Phi_{\rm G}$ is a Gaussian random field.
The non--Gaussianity parameter $f_{\rm NL}$ is often considered to be a
constant, yielding non--Gaussianities of the {\it local} type with a
bispectrum  which is maximized for squeezed configurations~\cite{Babich:2004gb}. 
The standard observables to constrain non--Gaussianities are the CMB and the Large-Scale Structure
(LSS) of the Universe. References \cite{Dalal:2007cu} and~\cite{Matarrese:2008nc} showed that primordial non--Gaussianities affect the clustering of dark matter halos inducing a scale-dependent 
large-scale bias~\cite{Slosar:2008hx,Afshordi:2008ru,Carbone:2008iz,Grossi:2009an,Desjacques:2008vf,
Pillepich:2008ka,Carbone:2010sb}. 

Following Refs.~\cite{Dalal:2007cu,Slosar:2008hx,Wands:2009ex} (see also
\cite{Bruni:2011ta,Yoo:2011zc,Jeong:2011as} for recent studies with
general relativity corrections), we consider a scale dependent bias
induced by the local non--Gaussianity  of the following form 
\begin{equation}
  \delta_g^S = b\, \delta_{\rm dm}^S \quad \mbox{where}\quad b=b_{\rm G}+\Delta b~,
\label{eq:deltaNG}
\end{equation}
 with $b_{\rm G}$, a constant Gaussian bias and 
\begin{equation}
\Delta b=3 f_{\rm NL}(1-b_{\rm G})\delta_{\rm c}\frac{H_0^2\Omega_{\rm m}}{k^2 T(k) D(a)}~.
  \label{eq:bfnl}
\end{equation}
$T(k)$ is the linear transfer function that we have taken to be equal to unity and $D(a)$ is the growth 
factor defined as $\delta_{\rm dm}(a)/\delta_{\rm dm}(a=1)$. The linear overdensity for spherical collapse 
can be considered as a constant: $\delta_{\rm c}=1.686$~\cite{Kitayama:1996pk}.

\begin{figure}[h!]
\vspace{-0.1cm}
\begin{center}
\begin{tabular}{cc}
\includegraphics[width=8cm]{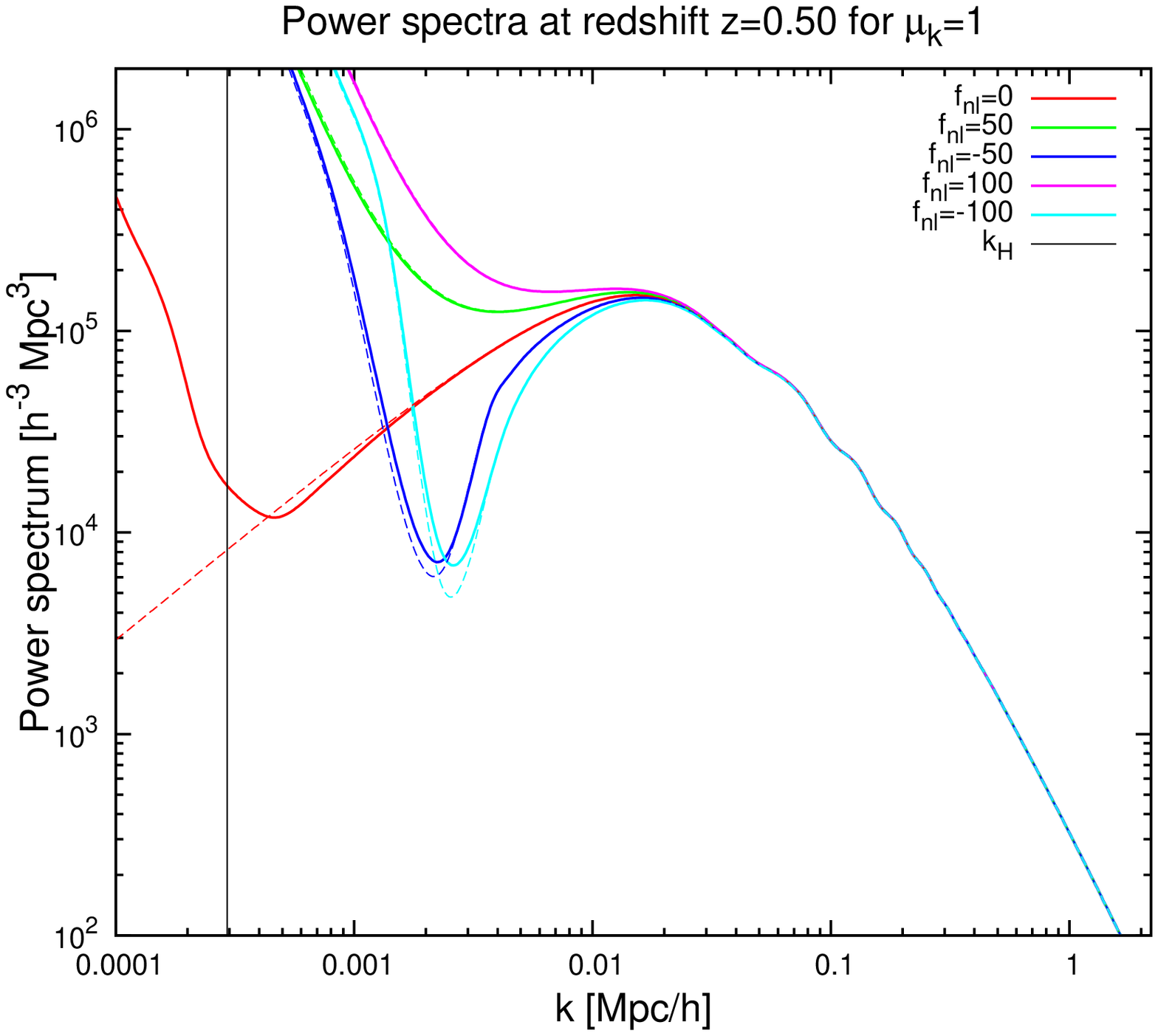}&\includegraphics[width=8cm]{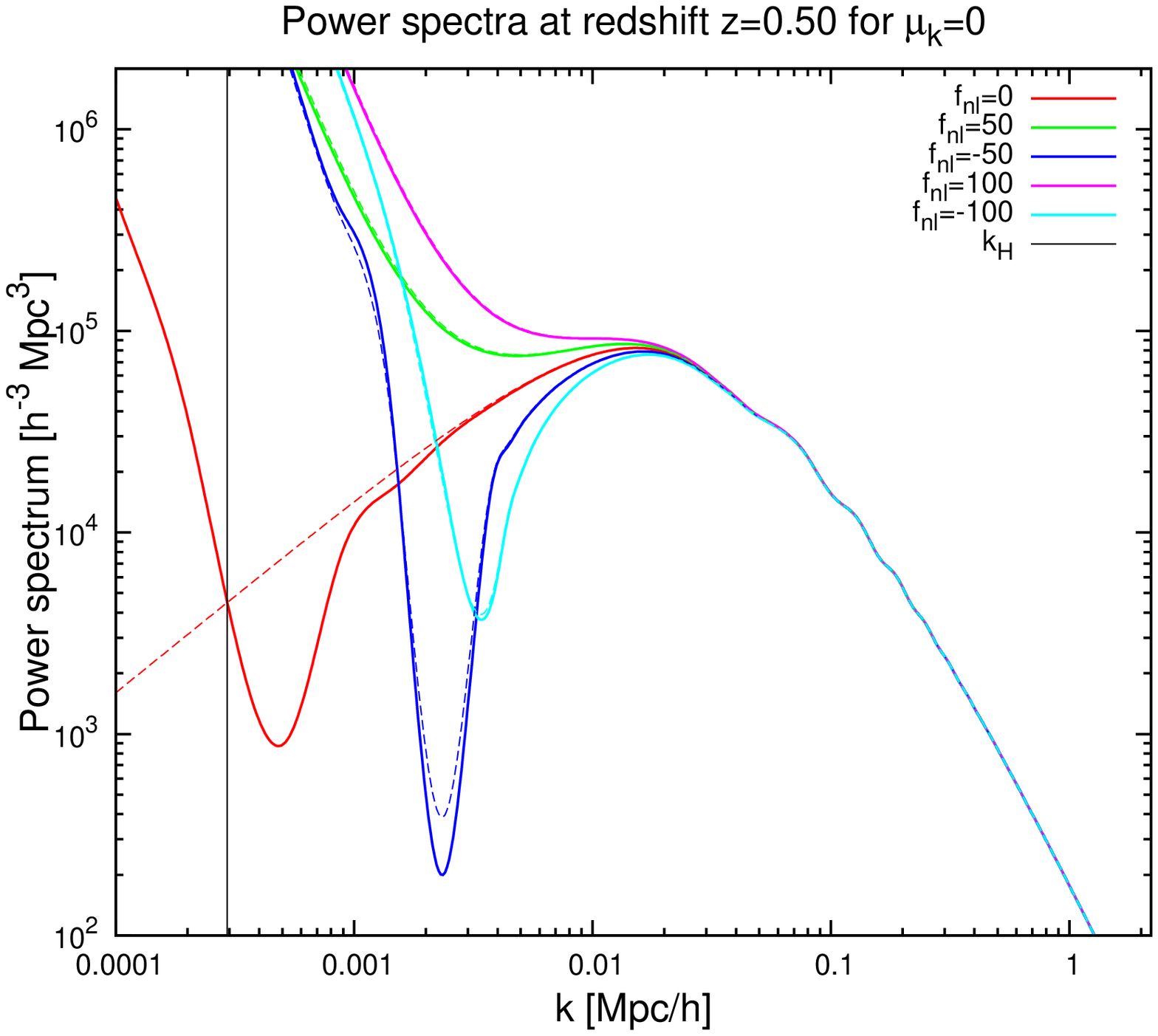}
\end{tabular}
\caption{\it \it $P_{\Delta_{obs}}(k)$ (solid lines) and  $P_{\Delta_{st}}(k)$
  (dashed lines) for $\mu_k=1$ (left) and $\mu_k=0$ (right) for
  five different values of the parameter $f_{\rm NL}$, for a Gaussian
  bias $b_{\rm G}=2$ and  $z=0.5$. The vertical lines depict the horizon scale $k_H$ for $w=-0.9$.}
\label{fig:Pfnl}
\end{center}
\end{figure}

The resulting non Gaussian halo power spectrum is shown in Fig.~\ref{fig:Pfnl}. 
The standard Newtonian power spectrum is now obtained using 
\begin{equation}
P_{\Delta_{st}}=P^S_{\rm dm}\left(b_G+\Delta b+f_{\rm eff}\mu_k^2\right)^2~,
\label{eq:Pstb}
\end{equation}
while the general relativity-corrected power spectra is obtained 
expressing Eq.~(\ref{eq:DeltaDur}) in the synchronous gauge and  
replacing $\delta_{\rm m}$ by the galaxy density fluctuation defined 
in Eq.~(\ref{eq:deltaNG}). The left (right) panel of Fig.~\ref{fig:Pfnl} 
shows the power spectra for the line of sight (transverse) modes. Note that
even in the absence of general relativity corrections the introduction of a negative
$f_{\rm NL}$ induces the presence of a dip at large scales (contrarily
to the case of positive $f_{\rm NL}$). This can be easily understood by
studying the $k$ dependence of the factor multiplying $P^S_{\rm dm}$ in
Eq.~(\ref{eq:Pstb}). In the negative $f_{\rm NL}$ case, once we
introduce general relativity corrections, the dip at large scales can
become shallower (deeper)  in the $\mu_k=1$ ($\mu_k=0$) case. In the case of
the positive $f_{\rm NL}$ values considered here, the presence of non 
Gaussianities induces an increase of the Newtonian matter power 
spectrum at $k< 0.01$~Mpc/$h$. General relativity corrections 
may also induce an increase of the power spectrum but at 
larger scales, $k< 0.001$~Mpc/$h$. However, non--Gaussianities
dominate the shape of the power spectrum and make 
the general relativity effects totally subdominant.
The shape of the non--Gaussian power spectrum barely changes when general 
relativistic effects are considered, regardless of the sign of the non 
Gaussianity parameter $f_{\rm NL}$. Therefore, we 
do not expect an important improvement on the precision measurement 
of the different 
 cosmological parameters nor large biases on them in a non--Gaussianity 
scenario when general relativity corrections are included, see
Sec.~\ref{sec:fisher} for a quantitative analysis.

\subsection{Coupled and modified gravity cosmologies}
\label{sec:coupl-modif-grav}
Interactions within the dark sectors, i.e. between cold dark matter and dark energy, are still allowed by 
observations~\cite{Amendola:1999er,Amendola:1999dr,Amendola:1999qq,Amendola:2000uh,Amendola:2003wa,
Amendola:2006dg,Valiviita:2008iv,He:2008si,Jackson:2009mz,Gavela:2009cy,CalderaCabral:2009ja,Valiviita:2009nu,
Majerotto:2009np,Gavela:2010tm,Honorez:2010rr,Martinelli:2010rt,LopezHonorez:2010ij}.
Constraints on coupled cosmologies as well as on modified gravity models could also be affected by the relativistic 
effects on the matter power spectrum. As an illustration, we parameterize the dark matter-dark 
energy interactions at the level of the stress-energy tensor conservation equations. Following the notations of 
\cite{Gavela:2010tm}, an energy momentum exchange of the following form can be introduced:
\begin{eqnarray}
\nabla_\mu T^\mu_{({\rm dm})\nu} =Q_\nu \quad\mbox{and}\quad
\nabla_\mu T^\mu_{({\rm de})\nu} =-Q_\nu~,
\label{eq:conservDMDE}
\end{eqnarray}
with
\begin{equation}
    Q_\nu= \xi {\mathcal H} \rho_{\rm de} u_{\nu}^{\rm dm}/a \qquad\mbox{or}\qquad  
    Q_\nu= \xi {\mathcal H} \rho_{\rm de} u_{\nu}^{\rm de}/a~,
\label{eq:ourm}
\end{equation}
where $ u_{\nu}^{\rm dm (de)}$ is the cold dark matter (dark energy) four velocity and $\xi$ is a dimensionless 
coupling, considered negative in order to avoid early time non adiabatic instabilities~\cite{Gavela:2009cy}.
In general, coupled models with $Q_\nu$ proportional to $u_{\nu}^{\rm de}$ are effectively modified gravity models. 
Assuming a flat universe and perfect measurements of $\Omega_{\rm dm} h^2$, $\Omega_{\rm b} h^2$, and of the angular diameter 
distance to the last scattering surface from Cosmic Microwave Background (CMB) observations~\cite{Komatsu:2010fb}, 
the amplitude of $\xi$ is degenerate with the physical energy density in dark matter today, $\Omega_{\rm dm} h^2$. Consequently, $\Omega_{\rm dm} h^2$ should be changed accordingly each time $\xi$ is varied, see Appendix B of 
Ref.~\cite{Honorez:2010rr} for the values of $\Omega_{\rm dm} h^2$ and $h$ considered here. 

Coupled cosmologies imply some extra terms in the expression of the gauge invariant matter fluctuation 
Eq.~(\ref{eq:Deltaz}). Indeed, in the case of the coupled models studied here
\begin{equation}
  \frac{d\rho_{ \rm dm}}{dz}=3\frac{\rho_{\rm  dm}}{1+z}-
  \xi\frac{\rho_{\rm  de}}{1+z}~,
  \label{eq:drhodz}
\end{equation}
which directly affects the expressions for $\Delta_m$ and $\Delta_z$. In the $u^{\rm dm}_{\nu}$ case 
the gauge invariant quantity defined in Eq.~(\ref{eq:rhom}) becomes:
\begin{equation}
\Delta_z^{u_{\nu}^{\rm dm}}= \Delta_{\rm dm}^\xi+
\left(3-\xi\frac{\rho_{\rm de}}{\rho_{\rm dm}}\right)\, \left[\mathbf{
    n \cdot V_{\rm dm} }+ \left(\Psi_B-\Phi_B\right) +\int_0^{r_s} d\lambda
\left(\dot \Psi_B-\dot \Phi_B\right) \right]~,
\end{equation}
where we have made explicit the $\xi$ dependence of the gauge invariant dark matter density perturbation:
\begin{equation}
\Delta_{\rm dm}^\xi= \delta_{\rm dm}+(3-\xi{\rho_{\rm de}}/{\rho_{\rm dm}}){\cal R}/{\cal H}~,
\end{equation}
where $\cal R$ is the curvature perturbation defined in Eq.~(\ref{eq:R}).
In the $u^{\rm de}_\nu$ case another extra contribution results from the modified Euler equation. 
In the gauge invariant formalism for dark matter perturbations (see e.g.~\cite{Gavela:2010tm})
\begin{equation}
  \dot {\mathbf V}_{\rm dm}=-{\cal H} {\mathbf V}_{\rm dm}-{\mathbf \nabla}\Psi_B + 
  \xi \frac{\rho_{\rm de}}{\rho_{\rm dm}} {\cal H}({\mathbf V}_{\rm de}-{\mathbf V}_{\rm dm})~.
\end{equation}
Therefore, the perturbation in the number density of galaxies in the
$u^{\rm de}_\nu$ case reads
\begin{equation}
  \Delta^{u_{\nu}^{\rm de}}_z= \Delta^{u_{\nu}^{\rm dm}}_z- \xi
  \frac{\rho_{\rm de}}{\rho_{\rm dm}} ({\mathbf V}_{\rm de}-{\mathbf V}_{\rm dm})\cdot
  {\mathbf n}~.
\end{equation}

\begin{figure}[tb!]
\vspace{-0.1cm}
\begin{center}
\begin{tabular}{cc}
  \includegraphics[width=8cm]{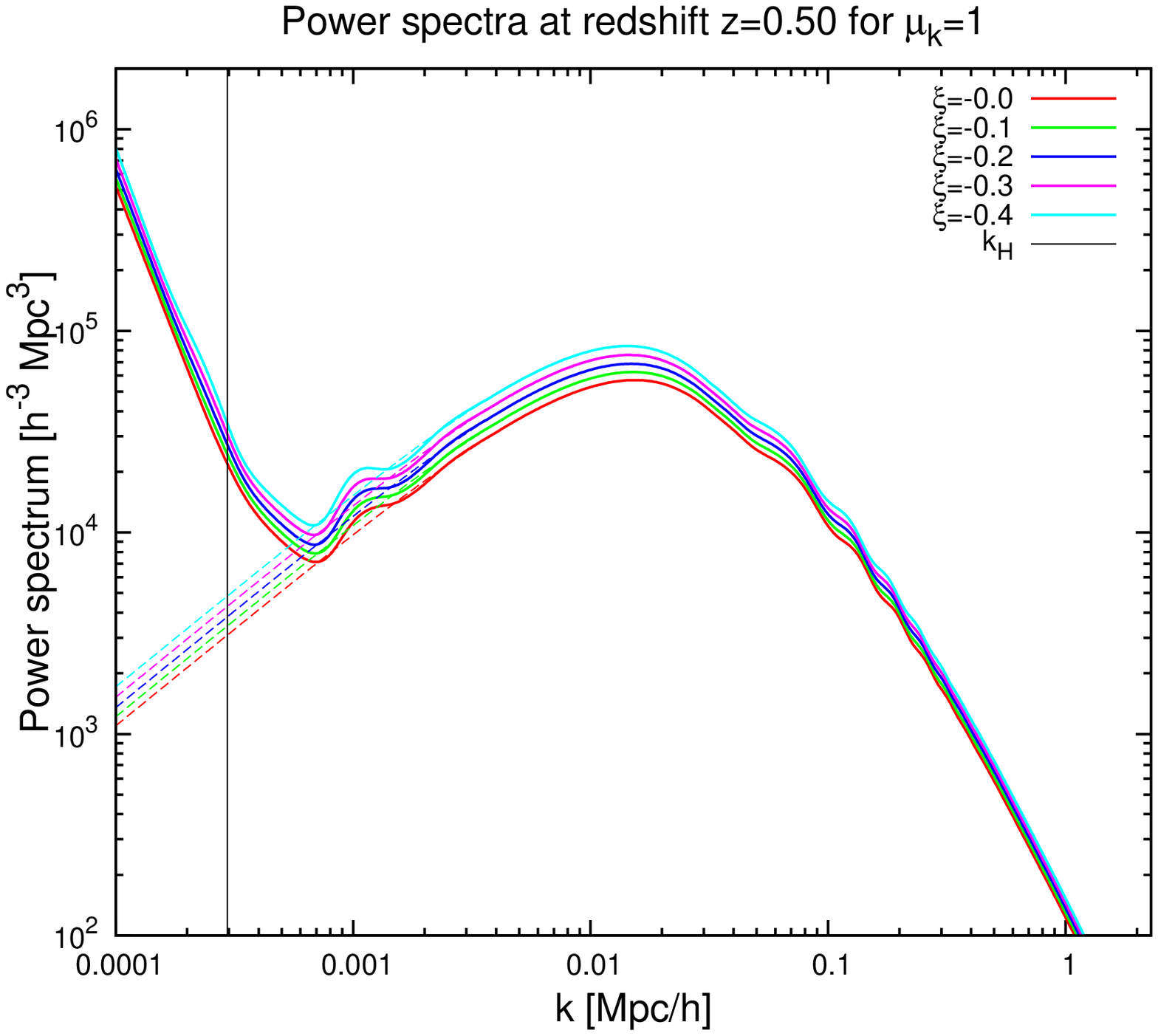}&\includegraphics[width=8cm]{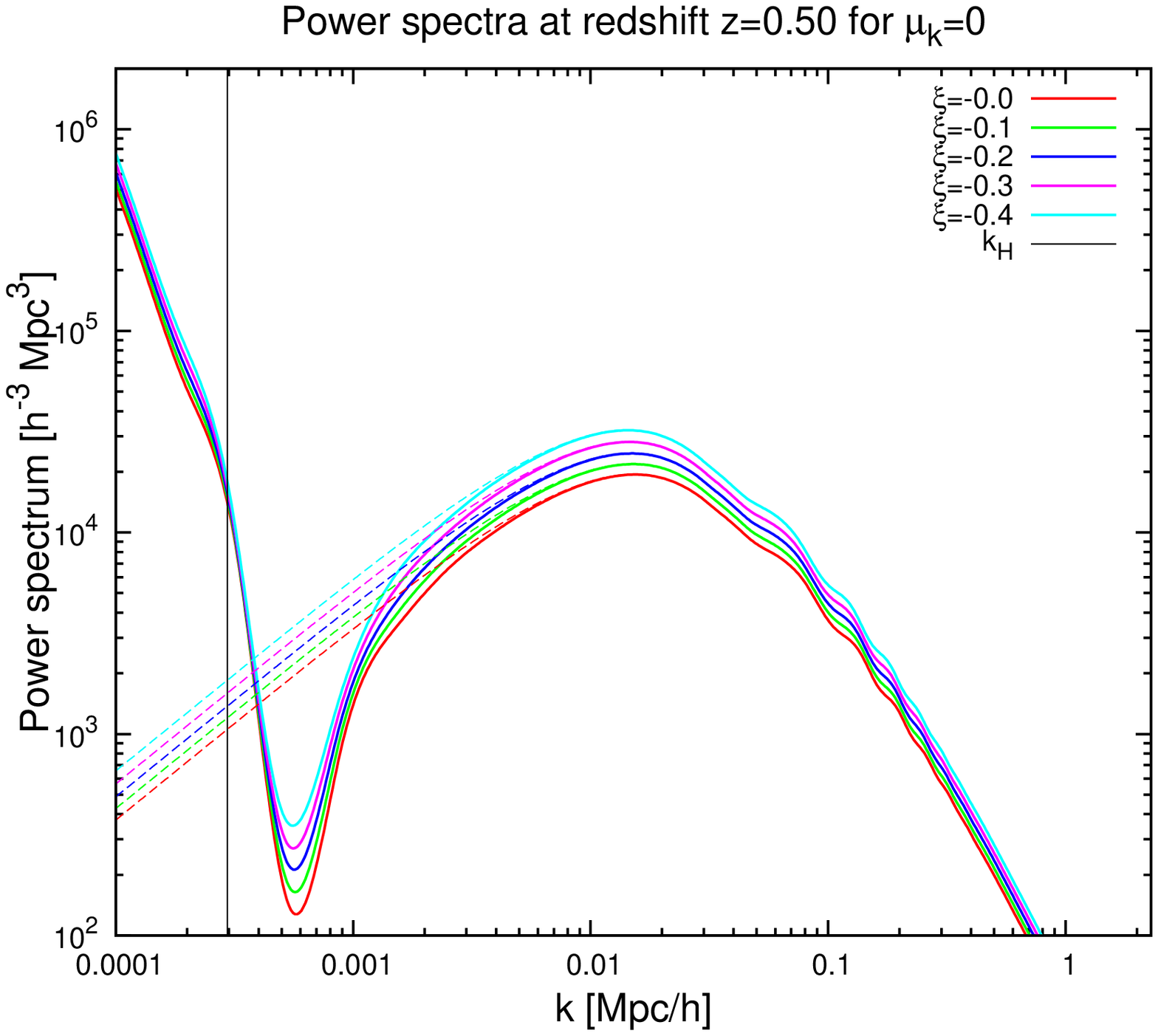}
\end{tabular}
\caption{\it The left (right) panel depicts $P_{\Delta_{obs}}(k)$ and $P_{\Delta_{st}}(k)$ by solid and dashed 
lines, respectively for coupled models $\propto u_{\nu}^{\rm dm}$ and $\mu_k=1$ ($\mu_k=0$). Different values of 
the coupling $\xi$ are illustrated, and the redshift is $z=0.5$. In all the models, the cosmological parameters 
$\Omega_{\rm dm} h^2$ and $h$ have been chosen to satisfy CMB constraints. The vertical lines show the horizon 
scale for $w=-0.9$.} 
 \label{fig:Pcoupl}
\end{center}
\end{figure}
Figure \ref{fig:Pcoupl} depicts the resulting matter power spectra $P_{\Delta_{obs}}(k,\mu_k)$ and 
$P_{\Delta_{st}}(k,\mu_k)$ for coupled models with an interaction term proportional to $u_{\nu}^{\rm dm}$ 
for both the line of sight and transverse modes at $z=0.5$ and for
different values of the coupling $\xi$. 
Notice that in coupled cosmological scenarios considered here, the
growth function appearing in the definition of
$P_{\Delta_{st}}(k,\mu_k)$ in  Eq.~(\ref{eq:Pst}) is given by
$d\ln\delta_{\rm dm}/d\ln a+\xi \rho_{ \rm de}/ \rho_{ \rm dm}$~\cite{Honorez:2010rr}. 
Similar results are obtained for the case in which the coupling term is proportional to $u_{\nu}^{\rm de}$. As in the case of the dark energy equation of state, no strong biases are expected 
in constraining the coupling when these new general relativistic terms are included in the analysis: 
the shape of the different curves including relativistic corrections barely changes when the coupling is varied.

\subsection{Neutrino masses}
\label{sec:neutrino-masses}

Consider a $\Lambda$CDM model plus massive neutrinos of a given energy density $\Omega_\nu h^2$. 
We would like to determine if the massive neutrino energy density could affect the position of the dip 
appearing in the dark matter power spectrum $P_{\Delta_{obs}}(k,\mu_k)$ for the transverse modes ($\mu_k=0$).
In order to simplify the discussion let us consider the expression of $\Delta_{obs}$ in the Newtonian gauge, 
see Eq.~(\ref{eq:DobsN}). In the approximation in which all projected quantities in the power spectrum 
computation are neglected, the dip appears for $\mu_k=0$ (i.e. ${\mathbf n} \cdot {\mathbf v}=0$) when the condition
\begin{equation}
\delta_{\rm dm}^N+\Psi_N-2\Phi_N +\frac{1}{\cal H} \dot \Phi_N+\left(\frac{2}{r_s{\cal H}}+
\frac{\dot {\cal H}}{{\cal H}^2}\right)\Psi_N=0
\label{eq:neut1}
\end{equation}
is satisfied. For a specific choice of redshift and of cosmology, the factor 
$\Sigma=({2}/{(r_s{\cal H})}+\dot {\cal H}/{{\cal H}^2})$ does not depend on the wave number $k$. 
Neglecting anisotropic stress, so that $\Psi_N =\Phi_N$, and making use of the Einstein equations 
(see Ref.~\cite{Ma:1995ey} for the prescription used here): 
\begin{eqnarray}
   &&k^2\Psi_N= -\frac32{\cal H}^2 \sum_a \Omega_a
   \left(\delta^N_a+3\frac{\cal H}{k} (1+w_a)v_a\right)\,,\\
  && k^2(\dot \Phi_N+{\cal H} \Psi_N)= \frac32{\cal H}^2 \sum_a
   \Omega_a(1+w_a)k v_a\,,
\end{eqnarray}
the dip position in the Fourier space as a function of $\Omega_a,\delta_a, v_a$ can be extracted: 
\begin{equation}
  k^2=\frac{3}{2\delta_{\rm dm}} {\cal H}^2\sum_a\Omega_a\left[(\Sigma-2)
  \delta_a+ (w_a+1) v_a\left(3\frac{\cal H}{k} (\Sigma-2)-\frac{k}{\cal H}\right)\right]~.
\end{equation}
In the previous equations the index $a$ runs over all the relevant fluids. In principle, different scenarios 
with different $\Omega_\nu$ will show a dip at different wave numbers. However, this $k$ difference will 
vanish when the total matter energy density (i.e. cold dark matter plus baryons plus the neutrino contribution) 
is kept constant. Therefore, general relativity effects can not help in extracting the values of the neutrino masses.

\section{Cosmological parameter forecasts and biases}
\label{sec:fisher}
In this section we explore if the measurement of the different cosmological parameters is affected by relativistic corrections. We present constraints from future galaxy survey measurements, 
making use of the Fisher matrix formalism. Then, we compare the cosmological parameter errors with and 
without general relativistic corrections. 
\subsection{Methodology}
\label{sec:methodology}
The Fisher matrix is defined as the expectation value of the second
derivative of the likelihood surface about the maximum. As long as the
posterior distribution for the parameters is well approximated by a
multivariate Gaussian function, its elements are given 
by~\cite{Tegmark:1996bz,Jungman:1995bz,Fisher:1935bi}
\begin{equation}
\label{eq:fish}
F_{\alpha\beta}=\frac{1}{2}{\rm
  Tr}\left[C^{-1}C_{,\alpha}C^{-1}C_{,\beta}\right]~,
\end{equation}
where $C=S+N$ is the total covariance which consists of signal
$S$ and noise $N$ terms. The commas in Eq.~(\ref{eq:fish})
denote derivatives with respect to the cosmological parameters
within the assumed fiducial cosmology. The 1--$\sigma$ error on a 
given parameter $p_\alpha$ marginalized over the other parameters is
$\sigma(p_\alpha)=\sqrt{({F}^{-1})_{\alpha\alpha}}$, ${F}^{-1}$ being
the inverse of the Fisher matrix. In order to focus on the role
played by general relativity corrections, we have restricted the analysis 
to galaxy survey data, i.e. we have not included in the analysis forecasts 
from the on going Planck CMB experiment. We exploit here an enlarged version 
of the future Euclid galaxy survey experiment, with an area of $20000$~deg$^2$, 
24 redshift slices between $z=0.15$ and $z=2.55$ and a mean galaxy density 
of $1.56 \times 10^{-3}$, see Refs.~\cite{Refregier:2006vt,Refregier:2010ss}.

Two possible fiducial cosmologies are analyzed: i) a constant 
$w$ cosmology ($w$ denotes the dark energy equation of state), and ii) a
constant $w$ cosmology with the presence of primordial non
Gaussianities (characterized by the parameter $f_{\rm NL}$). 
In the analysis i), the model is described by the physical baryon
and cold dark matter densities, $\Omega_{\rm b}h^2$ and
$\Omega_{\rm dm}h^2$, the scalar spectral index, $n_{\rm s}$, $h$, the
dimensionless amplitude of the primordial curvature perturbations,
$A_{\rm s}$ and  $w$. 
In the analysis ii), which includes non--Gaussianities, the model is 
described by $\Omega_{\rm b}h^2$, $\Omega_{\rm dm}h^2$, $h$, $w$, the
dark energy sound speed squared $c_{\rm s}^2$ and the $f_{\rm NL}$
parameter. We have therefore fixed in this case the scalar spectral
index and the dimensionless amplitude of primordial fluctuations,
expected to be measured with excellent accuracy by the CMB Planck
experiment. We follow a conservative approach,  
assuming that non--Gaussianities are
constrained exclusively from the very large scale halo power
spectrum.   

In addition to the marginalized parameter errors, 
the biases induced in the cosmological parameters when data are 
wrongly fitted to the standard Newtonian power spectrum, neglecting 
general relativity corrections, are also computed. The biases 
in the cosmological parameters read~\cite{dePutter:2007kf}
\begin{equation}
\delta p_\alpha =
(F^{-1})_{\alpha\beta} \sum_i \frac{\partial {\cal
    O}_{obs}^i}{\partial p_\beta} \frac{1}{\sigma^2_{{\cal O}^i_{obs}}} \left( {\cal O}^i_{obs}- {\cal O}^i_{st}\right)~,
\end{equation}
where the sum runs over the bins indices in $i= z$, $k$ and $\mu_k$ in the case
of the 3-D power spectrum analysis, i.e. ${\cal O}= P(z,k,\mu_k)$, and $i=
z$ and  $\ell$  in the case
of the 2-D power spectrum analysis , i.e. ${\cal O}= C_\ell(z)$.
  $F$ is the Fisher matrix computed with the power
spectra including general relativity corrections, ${\cal O}_{obs}^{(k,z)}$
and ${\cal O}_{st}^{(k,z)}$ are the general relativity and standard Newtonian
power spectra respectively and $\sigma_{{\cal O}^i_{obs}}$ is the error on the power spectrum with
general relativity corrections.  

In the case of analysis ii), we also have determined the shifts in the 
parameters $\{\Omega_{\rm  b}h^2,\Omega_{\rm dm}h^2,h,w,c_{\rm s}^2 \}$ that 
would result when mock data generated with primordial non--Gaussianities 
($f_{\rm NL}=20$ in this example) are fitted to a theoretical model without them.
The idea is the following: if the data are fitted assuming a model
$M_1$ with $n_1$ parameters, but the true underlying cosmology is a  
model $M_2$ characterized by $n_2$ parameters (with $n_2>n_1$ and the parameter space of $M_2$  
includes the model $M_1$ as a subset),
the inferred values of the $n_1$ parameters will be shifted from their  
true values to compensate for the fact that the model used to fit the  
data is wrong. In the case illustrated here, $M_2$ will
be the model \emph{with} non--Gaussianities and $M_1$ the one  
\emph{without} non--Gaussianities, i.e. with $f_{\rm NL}=0$.
While the first $n_1$ parameters are the same for both models, the  
remaining $n_2-n_1$ parameters in the enlarged model $M_2$ are accounting  
for the presence of non--Gaussianities, i.e. $f_{\rm NL}$.
Assuming a Gaussian likelihood, the shifts of the remaining $n_1$  
parameters are given by \cite{Heavens:2007ka}:
\begin{equation}
\delta\theta'_\alpha =
-(G^{-1})_{\alpha\beta}{F}_{\beta\zeta}\delta\psi_\zeta \qquad
\alpha,\beta=1\ldots n_1, \zeta=n_1+1\ldots n_2 \label{offset}~,
\end{equation}
where $G$ represents the Fisher sub-matrix for the model $M_1$ and $F$ denotes the Fisher matrix 
for the model $M_2$. In the case considered in this paper, $M_1$ is the model without  
primordial non--Gaussianities  while $f_{\rm NL}\neq 0$ in the model $M_2$ so that 
$n_2-n_1=1$ and $\delta \psi=\delta f_{\rm NL}=20$ .   

\subsection{3-D Power Spectrum}
\label{sec:3-d-power}
For details regarding the calculation of the Fisher matrix for the 3-D power spectra 
$P(k, \mu_k)$ measured by a galaxy survey, see Ref.~\cite{Seo:2003pu}. Here we perform 
a binning both in $k$ and in $\mu_k$, considering nine bins in the former quantity. 
The minimum scale $k_{min}$ is fixed to $10^{-4}$ $h$/Mpc and the
maximum scale is fixed to $0.1$ $h$/Mpc. 

Table~\ref{tab:w} contains the 1--$\sigma$ marginalized errors on the 
cosmological parameters for analysis i), with a fiducial  cosmology with constant dark energy equation 
of state $w=-1$. Two results are illustrated: those obtained
with the standard Newtonian power spectrum and those obtained with general
relativistic corrections included. 
 Note that the errors obtained in the standard Newtonian prescription
 are generally $40\%$ smaller than those obtained with
 general relativistic one, except for the $w$ parameter in which case 
the tendency is reversed.
The biases on the cosmological parameters are also presented
in Tab.~\ref{tab:w}. Note that their size is always smaller than the
1--$\sigma$ marginalized errors and therefore these biases will barely
interfere with the extraction of the cosmological parameters.   

Table~\ref{tab:fnl} presents the results from analysis ii), which includes non--Gaussianities 
with a fiducial $f_{\rm NL}=20$. Recently, the authors of Ref.~\cite{Yoo:2011zc} have shown 
that using methods to reduce the sampling variance and shot 
noise~\cite{Seljak:2008xr,Seljak:2009af,Hamaus:2011dq}, 
a full sky galaxy survey can measure general relativistic effects. 
We do not exploit here these cancellation methods, leaving these 
combined techniques for a future study.

The errors on cosmological parameters resulting from the
Fisher analysis are not improved including general relativity 
corrections. This fact was not unexpected, given that for the value
of the $f_{\rm NL}$ considered in this analysis the changes in the power
spectrum due to general relativity corrections are almost hidden by the effect
of non--Gaussianities, see Sec.~\ref{sec:non-gaussianity}. 
Note also that the biases are always smaller than the corresponding
1--$\sigma$ marginalized errors and therefore they will have no impact on the
extraction of the cosmological parameters. Also, we find no
significant shifts in the values of the cosmological parameters in any
of the two prescriptions when the non--Gaussianity parameter $f_{\rm NL}$ 
is (wrongly) assumed to be zero. We conclude that relativistic corrections in 
the 3-D power spectrum will not help in constraining the cosmological parameters.

\begin{table}[htbp]
\begin{center}
\begin{tabular}{cccc}
\hline\hline
 Parameter   &  $P_{\Delta_{st}}(k, \mu_k)$   & $P_{\Delta_{obs}}(k,\mu_k)$ &
Biases \\
\hline
$\Delta (\Omega_{\rm dm}h^2)$   &  $0.0035$ & $0.0057$ & $7.0
\,10^{-5}$ \\
$\Delta (\Omega_{\rm b}h^2)$   &  $0.0010$ & $0.0016$ &$-8.0
\,10^{-5}$\\
$\Delta A_{\rm s}$   &   $0.021$ & $0.036$ & $1.1 \,10^{-5}$\\
$\Delta h$ &  $0.010$ & $0.017$ &  $1.3 \,10^{-4}$\\
$\Delta n_{\rm s}$    &   $0.012$ & $0.016$ & $-4.3 \,10^{-3}$\\
$\Delta w$  &  $0.015$ &  $0.010$ &$7.7 \,10^{-3}$ \\
\hline\hline
\end{tabular}
\caption{1--$\sigma$ marginalized errors from the 
  Euclid-like survey considered here for a fiducial cosmology
  with a constant dark energy equation of state, with a fiducial value
  $w=-1$. The third row illustrates the biases induced in the
  cosmological parameters when general relativistic corrections are
  (wrongly) neglected.  The error on the amplitude of the primordial 
  fluctuations $\Delta A_{\rm s}$ is quoted in units of $2.64 \cdot 10^{-9}$.}
\label{tab:w}
\end{center}
\end{table}

\begin{figure}[tb!]
\vspace{-0.1cm}
\begin{center}
  \includegraphics[width=8cm]{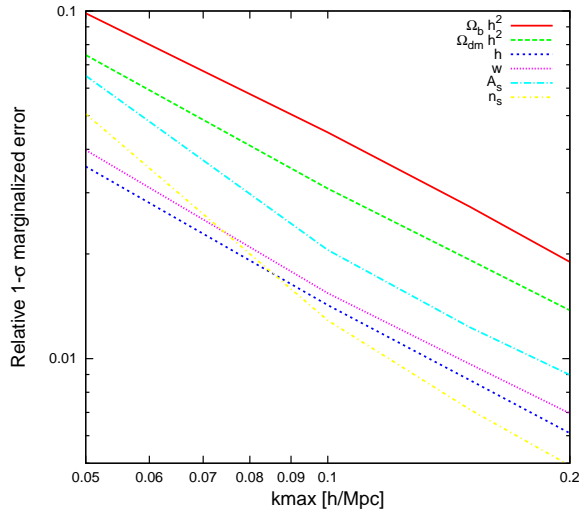}
\caption{\it Illustration of the relative 1--$\sigma$
  marginalized errors ($\Delta p / |p|$)  dependence on the scale
  $k_{max}$ using the standard Newtonian prescription for the cosmological parameters
  $p= \Omega_bh^2, \Omega_{dm}h^2, h, w, A_s$ and $n_s$ for a fiducial cosmology with a constant dark energy
  equation of state with a fiducial value $w=-1$. } 
 \label{fig:kmax}
\end{center}
\end{figure}

\begin{table}[htbp]
\begin{center}
\begin{tabular}{ccccc}
\hline\hline
 Parameter   &   $P_{\Delta_{st}}(k, \mu_k)$   &
 $P_{\Delta_{obs}}(k,\mu_k)$ &Biases &Shifts \\
\hline
$\Delta (\Omega_{\rm dm}h^2)$  & $6.2\, 10^{-4}$  & $6.1\, 10^{-4}$ & $-4.8\, 10^{-5}$&   $-2.4\, 10^{-4}$ \\
$\Delta (\Omega_{\rm b}h^2)$   & $8.7\, 10^{-4}$  & $9.3\, 10^{-4}$ & $-6.3\, 10^{-5}$&   $3.4\, 10^{-4}$ \\
$\Delta h$                    & $3.5\, 10^{-3}$  & $3.8\, 10^{-3}$ & $-3.5\, 10^{-4}$&   $2.0\, 10^{-3}$ \\
$\Delta w$                    & $1.3\, 10^{-2}$  & $2.0\, 10^{-3}$ & $-3.5\, 10^{-3}$&   $1.5\, 10^{-2}$ \\
$\Delta c_{\rm s}^2$           & 4.0  & 4.3 & 1.0 & -2.5 \\
$\Delta f_{\rm NL}$             & 3.1  & 3.1 & 0.7 & - \\
\hline\hline
\end{tabular}
\caption{1--$\sigma$ marginalized errors from the Euclid-like survey considered here 
 for a fiducial cosmology with a constant dark energy equation of state, 
 with fiducial values $w=-1$, $f_{\rm NL}=20$ and $c_{\rm s}^2=1$. The third row 
 presents the biases induced in the cosmological parameters when general 
 relativistic corrections are neglected. The shifts in the cosmological 
 parameters when $f_{\rm NL}$ is set to zero but the data are generated with 
 $f_{\rm NL}=20$ have been computed including general relativity corrections. 
 Similar results are obtained using the standard Newtonian expression.}
\label{tab:fnl}
\end{center}
\end{table}
Finally, we briefly comment on the dependence of the cosmological parameter errors on the maximum scale considered in the analysis, $k_{max}$, assuming a fiducial cosmology with a constant dark energy equation of state $w=-1$. A larger $k_{max}$ will imply a larger number of modes, more information from the location of the acoustic peaks is available and consequently the errors will be smaller. Figure~\ref{fig:kmax} illustrates the size of the relative errors on the different cosmological parameters considered in analysis i) versus the scale $k_{max}$. Going from $k_{max}= 0.05 h/$Mpc to $k_{max}= 0.2 h/$Mpc the expected errors in $\Omega_bh^2, \Omega_{dm}h^2, h $ and $w$ are reduced by a factor $\sim 5$ while in the case of  the $n_s$ parameter its error is reduced one order of magnitude. 

\subsection{2-D Angular Power Spectrum}
\label{sec:2-d-angular}
The 2-D $C_{\ell}$ angular power spectrum is a projection of the 3-D
quantity and therefore it implies an integration of the 3-D power
spectrum $P(k)$ convoluted with a window function, the Bessel
transform of the radial selection function, see
Refs.~\cite{Huterer:2000uj,Tegmark:2001xb}. Therefore, the 
$C_{\ell}$'s are not expected to give as much information
on the cosmological parameters as the 3-D power spectrum $P(k)$.  For
the calculations presented here we have computed the $C_{\ell}$
assuming no magnification bias and a constant distribution of sources 
with redshift. For details regarding the calculation of the Fisher
matrix for the 2-D power spectra measured by a galaxy survey, see
Ref.~\cite{Hu:2003pt} (notice that we considered $\ell_{min}=2$ and $\ell_{max}=400$).

Table~\ref{tab:wCl} contains the 1--$\sigma$ marginalized errors  for analysis i), a fiducial cosmology with constant dark energy equation
of state. We show the results when the Fisher matrix formalism is applied to 
the 2-D angular power
spectrum in the standard Newtonian case and
in the case in which general relativistic corrections are included. The errors in the two prescriptions are very similar. 
The biases in the cosmological parameters are also presented, and will
have very little impact in the measurement of the cosmological
parameters, as can be noticed from their sizes. 
 
Table~\ref{tab:fnlCl} presents the analogous but for the analysis ii)
with non--Gaussianities. Notice that the errors are exactly the
same for the two prescriptions and therefore there is no improvement
in the determination of the cosmological parameters when the general
relativistic corrections are addressed in the angular power spectrum.  
The biases induced in the cosmological parameters when the data are fitted
to the standard Newtonian power spectrum are also presented. These
biases are always smaller than the corresponding 1--$\sigma$
marginalized errors and therefore, will have no impact on the
extraction of the cosmological parameters. Also, we find no
significant shifts in the values of the cosmological parameters in any
of the two prescriptions when the non--Gaussianity parameter $f_{\rm NL}$ 
is (wrongly) assumed to be zero. Consequently, from what regards 
the 2-D angular power spectrum, relativistic corrections will
not have any impact on future measurements of the cosmological
parameters, even if the information contained at the largest scales 
becomes at reach.  

Notice that, as expected, the errors on the cosmological parameters
obtained exploiting the 2-D power spectrum are, in general, larger
than in the 3-D case. In the case of analysis i), the expected errors on w differ 
by one order of magnitude. The results of the 2-D and 3-D
analysis should however roughly match in the limit of many narrow redshift
bins.  We have thus carried out a new fisher matrix analysis, 
decreasing the size of the redshift bin one order of magnitude in the 2-D analysis. 
In the latter case, similar marginalized errors are obtained when exploiting 2-D or 
3-D power spectrum. In the case of the non--Gaussianity parameter $f_{\rm NL}$ the 
errors are three orders of
magnitude larger when using the 2-D angular information. This is due
to the fact that the 2-D $C_\ell$ angular power spectrum is
essentially sensitive to modes transverse to the line of sight, while
the  3-D $P(k)$ power spectrum benefit from extra information from the
radial modes.

\begin{table}[htbp]
\begin{center}
\begin{tabular}{cccc}
\hline\hline
 Parameter   &   $C_{\ell\,\Delta_{st}}$   & $C_{\ell\,\Delta_{obs}}$& Biases   \\
\hline
$\Delta (\Omega_{\rm dm}h^2)$  & $0.0088$& $0.0086$&$0.003$\\
$\Delta (\Omega_{\rm b}h^2)$   &$0.002$& $0.002$&$<10^{-4}$\\
$\Delta A_{\rm s}$   & $0.093$& $0.093$&$-0.03$\\
$\Delta h$ & $0.045$&$0.045$&$0.003$\\
$\Delta n_{\rm s}$  & $0.04$&$0.04$&$-0.02$\\
$\Delta w$  &  $0.24$ &$0.24$& $0.01$\\
\hline\hline
\end{tabular}
\caption{1--$\sigma$ marginalized errors from the Euclid-like survey
  considered here for a fiducial cosmology with a constant dark energy
  equation of state with a fiducial value $w=-1$. The biases in the
  parameters are also presented. The error on the
amplitude of the primordial fluctuations $\Delta A_{\rm s}$  is quoted in units of
$2.64 \cdot 10^{-9}$.}
\label{tab:wCl}
\end{center}
\end{table}

\begin{table}[htbp]
\begin{center}
\begin{tabular}{ccccc}
\hline\hline
 Parameter   &   $C_{\ell\,\Delta_{st}}$   & $C_{\ell\,\Delta_{obs}}$
 &Biases  &Shifts \\
\hline
$\Delta (\Omega_{\rm dm}h^2)$  & $0.0106$  &   $0.0106$  & $-0.0001$& $-0.00001$\\
$\Delta (\Omega_{\rm b}h^2)$  & $0.0053$  &   $0.0053$  & $-0.0001$& $< 10^{-6}$\\
$\Delta h$               & $0.0499$   &$0.0499$   & $-0.00004$&  $<10^{-5}$ \\
$\Delta w$               & $0.2276$   &$0.2264$   & $0.0006$& $-0.0007$\\
$\Delta c_{\rm s}^2$           & $4.886$    & $4.771$    &$-0.1527$& $0.025$\\
$\Delta f_{\rm NL}$          & $1870$     & $1688$     & $-1189$ &-\\
\hline\hline
\end{tabular}
\caption{1--$\sigma$ marginalized errors from the Euclid-like survey data 
 for a fiducial cosmology with a constant dark energy equation of state, 
 with fiducial values $w=-1$, $f_{\rm NL}=20$, $c_{\rm s}^2=1$. The third 
 row presents the biases induced in the cosmological parameters when general 
 relativistic corrections are neglected. The shifts in the cosmological 
 parameters when $f_{\rm NL}$ is set to zero but the data are generated with 
 $f_{\rm NL}=20$ have been computed including general relativity corrections. 
 Similar results are obtained using the standard Newtonian expression.}
\label{tab:fnlCl}
\end{center}
\end{table}

\section{Summary}
\label{sec:concl}

The complete general relativistic description of the observed matter power spectrum at large scales 
is significantly different than the standard Newtonian one. The observed redshift and position of galaxies
are  affected by the different matter fluctuations and by the gravity waves between the source 
galaxies and the observer, see Refs.~\cite{Yoo:2009au,Yoo:2010ni,Bonvin:2011bg,Challinor:2011bk}.  
In this paper we have studied the role of relativistic effects in the extraction of different cosmological 
parameters with the galaxy power spectrum measurements that will be available from future surveys.

We have explored the impact of such corrections in several cosmological scenarios as: constant (but $w\neq1$) 
dark energy equation of state, time varying $w(a)=w_0+w_a(1-a)$ dark energy equation of state, coupled dark 
matter dark energy scenario, massive neutrinos and primordial non--Gaussianities. We have performed a Fisher 
matrix analysis considering data from a future Euclid--like
spectroscopic galaxy survey for two scenarios: one with a constant
dark energy equation of state, the other with non--Gaussianities. 
We find that general relativistic corrections will not interfere
neither with the extraction of the standard 
cosmological parameters (as the cold dark matter and baryon densities) nor with the measurement of primordial 
non--Gaussianities. The expected marginalized errors when relativistic corrections are included in the matter 
or halo power spectra are very similar to those obtained in the standard Newtonian case. The biases induced 
in the different cosmological parameters when neglecting these relativistic effects are also negligible. 
We conclude that the measurement of the cosmological parameters will not be compromised by the presence 
of general relativistic effects, once they will be included in the analysis.

\section{Acknowledgments}
The authors would like to acknowledge R. de Putter, C. Pe\~{n}a Garay and L. Verde for very useful comments 
and enlightening discussions. O.M. is supported by AYA2008-03531 and the Consolider Ingenio-2010 project CSD2007-00060. 
L. L. H is supported in part by the IISN and by the Belgian Science Policy (IAP VI/11). 
S.~Rigolin acknowledges the partial support of an Excellence Grant of Fondazione Cariparo and of the 
European Program‚ Unification in the LHC era‚ under the contract PITN- GA-2009-237920 (UNILHC).

\appendix
\section{Gauge invariant formalism }
\label{sec:gauge-invar-form}
The conventions we use are from Ref.~\cite{Kodama:1985bj} with a few exceptions. More details can be found in  
Ref.~\cite{Gavela:2010tm}. For perturbations in a flat space time, the perturbation variables can be expanded 
by harmonic functions $Y^{(S)}(x,k)$ satisfying $(\nabla_x+k^2)Y^{(S)}=0$.
In the following we focus on scalar perturbations, for which we define:
\begin{eqnarray}
Y^{(S)} _i & = & - \frac{1}{k} Y^{(S)}_{|i}~,  \\
Y^{(S)} _{ij} & = & \frac{1}{k^2} Y^{(S)}_{|ij} + \frac{1}{3}
\gamma_{ij} Y^{(S)}~.  
\end{eqnarray}

Following Ref.~\cite{Kodama:1985bj}, the FRW metric, up to first order in perturbation 
theory, can be written as:
\begin{eqnarray}
g_{\mu \nu} dx^\mu dx^\nu &=&
 \bar{a}^2 \left[- (1+2A)d\tau^2 - 2 B_i d\tau dx^i + (\gamma_{ij} +2 H_{ij})dx^i dx^j \right]\,,
\label{metric}
\end{eqnarray}
where  $\gamma_{ij}$ is the 3D flat metric with positive signature. The perturbations $A$, $B_i$ and $H_{ij}$ are 
functions of time and space and are in general gauge-dependent. Expanding the independent perturbations in the 
Fourier basis, and keeping only the scalar modes, we denote:
\begin{eqnarray}
A       & \raw & \widetilde{A} \YS~; \nn \\
B_i     & \raw & \widetilde{B} \YSv~;  \nn \\
H_{ij}  & \raw & \widetilde{H}_L \gt \YS~ + \widetilde{H}_T \YSt \nn \,.
\end{eqnarray}
In the following, for the sake of simplicity, we will omit the tilde symbols 
in the notation. Remember that all these quantities are represented by the correspondent Fourier expansion and depend 
only on time and on the 3-momentum $k$, while the position dependence is left only in the $Y$ basis elements.

Using these metric perturbations, we can now define $\sigma_g$, the
shear perturbation and ${\mathcal{R}}$, the curvature perturbation, as
\begin{eqnarray}
\sigma_g &=&\frac{1}{k}\left(\dot{H}_T - k B\right)~; \label{eq:sg}\\
{\mathcal{R}} &=& H_L + \frac{1}{3} H_T~, \label{eq:R}
\end{eqnarray}
which are no gauge invariant quantities. The Bardeen metric gauge invariants are defined as~\cite{Bardeen:1980kt}: 
\begin{eqnarray}
\Psi_B &=& A - \frac{\Hu}{k} \sigma_g - \frac{1}{k} \dot\sigma_g \label{eq:bardeenpsi}~,  \\
\Phi_B &=& H_L+\frac{1}{3}H_T - \frac{\Hu}{k} \sigma_g~.
\label{eq:bardeenphi} 
\end{eqnarray}
In the same line one can define perturbations for the energy--density
for a given fluid $a$:
\begin{eqnarray}
u^\mu_a &=& \frac{1}{\bar{a}} \left(1-A,v^i_a\right)~;\\
T_a^{\mu \nu} &=& \bar{\rho}_a \left(1 + \delta_a \right) u^\mu_a u^\nu_a + \tau^{\mu \nu}~,
\end{eqnarray}
where $v^i_a$ is the peculiar velocity perturbation of the fluid and
$\delta_a$ the fluid matter density contrast. 
Following \cite{Kodama:1985bj} one define the following gauge-invariant quantities:
\begin{eqnarray}
V_a &=& v_a- \frac{\dot{H}_T}{k}~; \\
\Delta_a &=&\delta_a-\frac{\dot{\bar\rho}_a}{\bar{\rho}_a}\frac{\cal R}{\cal H}\, ,
\end{eqnarray}
where $\Delta_a$ is  the gauge invariant density contrast for the fluid
$a$ defined in the gravity rest frame. Notice that  $v_a^i= v_a Y^{i}$
and that in Eq.~(\ref{eq:Deltaz}) and the following, $V^i$ refers to the gauge
invariant velocity perturbation associated to the matter component, i.e.
$V^i\equiv V_{\rm m}^i=V_{\rm m} Y^i$.

\subsection{Photon wave vector: some relations} 
\label{sec:photon-wave-vector}
Here we provide several relations resulting from the null energy
condition $K^\mu K_\mu =0$ and  the geodesic equations $K^\mu
K^\nu_{;\mu} =0$ useful in the derivation of the expression of gauge
invariant matter density perturbation $\Delta_z$ defined in
Eq.~(\ref{eq:rhom}).   On the one hand, from the perturbed null equation  $K^\mu K_\mu
=0$, one obtains the following relation between  
the temporal and spatial null vector perturbations:
\begin{eqnarray}
n^i \delta n_i = \frac{\delta \nu}{\nu} + \left(\Psi_B -\Phi_B \right) - \frac{1}{k^2} \frac{d}{d \lambda}
\left( \frac{d H_T}{d \lambda} - 2 H_T + k B \right) \, ,
\end{eqnarray}
where $d/d\lambda =\partial_\tau+n^i\partial_i$ and we have taken into account that the background 
null equation imposes $n^i n_i =1$. We have also used the background geodesic equation giving rise to
$n_i \partial_j n^i = n_i \dot{n}^i = 0$. On the other hand, the temporal geodesic equation 
$K^\nu K^0_{;\nu}$ gives the following condition:
\begin{eqnarray}
\frac{d}{d\lambda} \left( \frac{\delta \nu}{\nu} + 2 \Psi_B \right) = \left(\dot\Psi_B -\dot\Phi_B \right) - 
\frac{1}{k} \frac{d}{d\lambda} \left( \frac{d \sigma_g}{d\lambda} +2 \Hu \sigma_g \right) \, .
\end{eqnarray}
\subsection{Newtonian gauge}
\label{sec:newtonian-gauge}
It can be useful for comparison to make a particular gauge choice. In the  Newtonian gauge, $\sigma_g=0$ and
the perturbed metric is reduced to:
\begin{equation}
  ds^2=a^2[-(1+2\Psi_N)d\tau^2+(1-2\Phi_N)dx^idx_i]~.
\label{eq:Newt}
\end{equation}
In particular, for this gauge choice metric perturbations are given by:
\begin{eqnarray}
  \Psi_N&=&\Psi_B=A~,\cr
  \Phi_N&=&-\Phi_B=-{\cal R}~.
\end{eqnarray}
The convergence $\kappa$, in the Newtonian gauge, reads:
\begin{equation}
  \kappa=\int_0^{r_s}\frac{r_s-r}{2r_s r} \Delta_\Omega(\Phi_N+\Psi_N)~,
\label{eq:kappa}
\end{equation}
where $r_s$ is the comoving distance between the source and the
observer and $\Delta_\Omega=\cot\theta \, \partial_\theta+\partial_\theta^2+1/\sin\theta^2\partial_{\phi}^2$ 
is the angular laplacian on a unit sphere.  
\bibliography{bibdmde-v2}{}

\begin{thebibliography}{10}

\bibitem{Bonvin:2005ps}
Camille Bonvin, Ruth Durrer, and M.~Alice Gasparini.
\newblock {Fluctuations of the luminosity distance}.
\newblock {\em Phys. Rev.}, D73:023523, 2006, astro-ph/0511183.

\bibitem{Yoo:2009au}
Jaiyul Yoo, A.~Liam Fitzpatrick, and Matias Zaldarriaga.
\newblock {A New Perspective on Galaxy Clustering as a Cosmological Probe:
  General Relativistic Effects}.
\newblock {\em Phys. Rev.}, D80:083514, 2009, 0907.0707.

\bibitem{Yoo:2010ni}
Jaiyul Yoo.
\newblock {General Relativistic Description of the Observed Galaxy Power
  Spectrum: Do We Understand What We Measure?}
\newblock {\em Phys. Rev.}, D82:083508, 2010, 1009.3021.

\bibitem{Bonvin:2011bg}
Camille Bonvin and Ruth Durrer.
\newblock {What galaxy surveys really measure}.
\newblock 2011, 1105.5280.

\bibitem{Challinor:2011bk}
Anthony Challinor and Antony Lewis.
\newblock {The linear power spectrum of observed source number counts}.
\newblock 2011, 1105.5292.

\bibitem{Jeong:2011as}
Donghui Jeong, Fabian Schmidt, and Christopher~M. Hirata.
\newblock {Large-scale clustering of galaxies in general relativity}.
\newblock 2011, 1107.5427.

\bibitem{Mandelbaum:2005nx}
Rachel Mandelbaum, Uros Seljak, Guinevere Kauffmann, Christopher~M. Hirata, and
  Jonathan Brinkmann.
\newblock {Galaxy halo masses and satellite fractions from galaxy-galaxy
  lensing in the sdss: stellar mass, luminosity, morphology, and environment
  dependencies}.
\newblock {\em Mon.Not.Roy.Astron.Soc.}, 368:715, 2006, astro-ph/0511164.

\bibitem{camb}
\url{http://camb.info/sources/}.

\bibitem{Chevallier:2000qy}
Michel Chevallier and David Polarski.
\newblock {Accelerating universes with scaling dark matter}.
\newblock {\em Int. J. Mod. Phys.}, D10:213--224, 2001, gr-qc/0009008.

\bibitem{Linder:2002et}
Eric~V. Linder.
\newblock {Exploring the expansion history of the universe}.
\newblock {\em Phys.Rev.Lett.}, 90:091301, 2003, astro-ph/0208512.

\bibitem{Albrecht:2006um}
Andreas Albrecht, Gary Bernstein, Robert Cahn, Wendy~L. Freedman, Jacqueline
  Hewitt, et~al.
\newblock {Report of the Dark Energy Task Force}.
\newblock 2006, astro-ph/0609591.

\bibitem{Linder:2006sv}
Eric~V. Linder.
\newblock {The paths of quintessence}.
\newblock {\em Phys.Rev.}, D73:063010, 2006, astro-ph/0601052.

\bibitem{Hu:2007pj}
Wayne Hu and Ignacy Sawicki.
\newblock {A Parameterized Post-Friedmann Framework for Modified Gravity}.
\newblock {\em Phys.Rev.}, D76:104043, 2007, 0708.1190.

\bibitem{Hu:2008zd}
Wayne Hu.
\newblock {Parametrized Post-Friedmann Signatures of Acceleration in the CMB}.
\newblock {\em Phys.Rev.}, D77:103524, 2008, 0801.2433.

\bibitem{Fang:2008sn}
Wenjuan Fang, Wayne Hu, and Antony Lewis.
\newblock {Crossing the Phantom Divide with Parameterized Post- Friedmann Dark
  Energy}.
\newblock {\em Phys. Rev.}, D78:087303, 2008, 0808.3125.

\bibitem{Bruni:2011ta}
Marco Bruni et~al.
\newblock {Disentangling non-Gaussianity, bias and GR effects in the galaxy
  distribution}.
\newblock 2011, 1106.3999.

\bibitem{Salopek:1990jq}
D.S. Salopek and J.R. Bond.
\newblock {Nonlinear evolution of long wavelength metric fluctuations in
  inflationary models}.
\newblock {\em Phys.Rev.}, D42:3936--3962, 1990.

\bibitem{Gangui:1993tt}
Alejandro Gangui, Francesco Lucchin, Sabino Matarrese, and Silvia Mollerach.
\newblock {The Three point correlation function of the cosmic microwave
  background in inflationary models}.
\newblock {\em Astrophys.J.}, 430:447--457, 1994, astro-ph/9312033.

\bibitem{Verde:1999ij}
Licia Verde, Li-Min Wang, Alan Heavens, and Marc Kamionkowski.
\newblock {Large scale structure, the cosmic microwave background, and
  primordial non-gaussianity}.
\newblock {\em Mon.Not.Roy.Astron.Soc.}, 313:L141--L147, 2000,
  astro-ph/9906301.

\bibitem{Komatsu:2001rj}
Eiichiro Komatsu and David~N. Spergel.
\newblock {Acoustic signatures in the primary microwave background bispectrum}.
\newblock {\em Phys.Rev.}, D63:063002, 2001, astro-ph/0005036.

\bibitem{Babich:2004gb}
Daniel Babich, Paolo Creminelli, and Matias Zaldarriaga.
\newblock {The Shape of non-Gaussianities}.
\newblock {\em JCAP}, 0408:009, 2004, astro-ph/0405356.

\bibitem{Dalal:2007cu}
Neal Dalal, Olivier Dore, Dragan Huterer, and Alexander Shirokov.
\newblock {The imprints of primordial non-gaussianities on large-scale
  structure: scale dependent bias and abundance of virialized objects}.
\newblock {\em Phys.Rev.}, D77:123514, 2008, 0710.4560.

\bibitem{Matarrese:2008nc}
Sabino Matarrese and Licia Verde.
\newblock {The effect of primordial non-Gaussianity on halo bias}.
\newblock {\em Astrophys.J.}, 677:L77--L80, 2008, 0801.4826.

\bibitem{Slosar:2008hx}
Anze Slosar, Christopher Hirata, Uros Seljak, Shirley Ho, and Nikhil
  Padmanabhan.
\newblock {Constraints on local primordial non-Gaussianity from large scale
  structure}.
\newblock {\em JCAP}, 0808:031, 2008, 0805.3580.

\bibitem{Afshordi:2008ru}
Niayesh Afshordi and Andrew~J. Tolley.
\newblock {Primordial non-gaussianity, statistics of collapsed objects, and the
  Integrated Sachs-Wolfe effect}.
\newblock {\em Phys.Rev.}, D78:123507, 2008, 0806.1046.

\bibitem{Carbone:2008iz}
Carmelita Carbone, Licia Verde, and Sabino Matarrese.
\newblock {Non-Gaussian halo bias and future galaxy surveys}.
\newblock {\em Astrophys.J.}, 684:L1--L4, 2008, 0806.1950.

\bibitem{Grossi:2009an}
M.~Grossi, L.~Verde, C.~Carbone, K.~Dolag, E.~Branchini, et~al.
\newblock {Large-scale non-Gaussian mass function and halo bias: tests on
  N-body simulations}.
\newblock {\em Mon.Not.Roy.Astron.Soc.}, 398:321--332, 2009, 0902.2013.

\bibitem{Desjacques:2008vf}
Vincent Desjacques, Uros Seljak, and Ilian Iliev.
\newblock {Scale-dependent bias induced by local non-Gaussianity: A comparison
  to N-body simulations}.
\newblock 2008, 0811.2748.

\bibitem{Pillepich:2008ka}
Annalisa Pillepich, Cristiano Porciani, and Oliver Hahn.
\newblock {Universal halo mass function and scale-dependent bias from N-body
  simulations with non-Gaussian initial conditions}.
\newblock 2008, 0811.4176.

\bibitem{Carbone:2010sb}
Carmelita Carbone, Olga Mena, and Licia Verde.
\newblock {Cosmological Parameters Degeneracies and Non-Gaussian Halo Bias}.
\newblock {\em JCAP}, 1007:020, 2010, 1003.0456.

\bibitem{Wands:2009ex}
David Wands and Anze Slosar.
\newblock {Scale-dependent bias from primordial non-Gaussianity in general
  relativity}.
\newblock {\em Phys.Rev.}, D79:123507, 2009, 0902.1084.

\bibitem{Yoo:2011zc}
Jaiyul Yoo, Nico Hamaus, Uros Seljak, and Matias Zaldarriaga.
\newblock {Testing General Relativity on Horizon Scales and the Primordial
  non-Gaussianity}.
\newblock 2011, 1109.0998.

\bibitem{Kitayama:1996pk}
Tetsu Kitayama and Yasushi Suto.
\newblock {Formation rate of gravitational structures and the cosmic x-ray
  background radiation}.
\newblock {\em Mon.Not.Roy.Astron.Soc.}, 280:638, 1996, astro-ph/9602076.

\bibitem{Amendola:1999er}
Luca Amendola.
\newblock {Coupled quintessence}.
\newblock {\em Phys. Rev.}, D62:043511, 2000, astro-ph/9908023.

\bibitem{Amendola:1999dr}
Luca Amendola.
\newblock {Perturbations in a coupled scalar field cosmology}.
\newblock {\em Mon. Not. Roy. Astron. Soc.}, 312:521, 2000, astro-ph/9906073.

\bibitem{Amendola:1999qq}
Luca Amendola.
\newblock {Scaling solutions in general non-minimal coupling theories}.
\newblock {\em Phys. Rev.}, D60:043501, 1999, astro-ph/9904120.

\bibitem{Amendola:2000uh}
Luca Amendola and Domenico Tocchini-Valentini.
\newblock {Stationary dark energy: the present universe as a global attractor}.
\newblock {\em Phys. Rev.}, D64:043509, 2001, astro-ph/0011243.

\bibitem{Amendola:2003wa}
Luca Amendola.
\newblock Linear and non-linear perturbations in dark energy models.
\newblock {\em Phys. Rev.}, D69:103524, 2004, astro-ph/0311175.

\bibitem{Amendola:2006dg}
Luca Amendola, Gabriela Camargo~Campos, and Rogerio Rosenfeld.
\newblock Consequences of dark matter - dark energy interaction on cosmological
  parameters derived from snia data.
\newblock {\em Phys. Rev.}, D75:083506, 2007, astro-ph/0610806.

\bibitem{Valiviita:2008iv}
Jussi Valiviita, Elisabetta Majerotto, and Roy Maartens.
\newblock {Instability in interacting dark energy and dark matter fluids}.
\newblock {\em JCAP}, 0807:020, 2008, 0804.0232.

\bibitem{He:2008si}
Jian-Hua He, Bin Wang, and Elcio Abdalla.
\newblock {Stability of the curvature perturbation in dark sectors' mutual
  interacting models}.
\newblock {\em Phys. Lett.}, B671:139--145, 2009, 0807.3471.

\bibitem{Jackson:2009mz}
Brendan~M. Jackson, Andy Taylor, and Arjun Berera.
\newblock {On the large-scale instability in interacting dark energy and dark
  matter fluids}.
\newblock {\em Phys. Rev.}, D79:043526, 2009, 0901.3272.

\bibitem{Gavela:2009cy}
M.~B. Gavela, D.~Hernandez, L.~Lopez Honorez, O.~Mena, and S.~Rigolin.
\newblock {Dark coupling}.
\newblock {\em JCAP}, 0907:034, 2009, 0901.1611.

\bibitem{CalderaCabral:2009ja}
Gabriela Caldera-Cabral, Roy Maartens, and Bjoern~Malte Schaefer.
\newblock {The Growth of Structure in Interacting Dark Energy Models}.
\newblock {\em JCAP}, 0907:027, 2009, 0905.0492.

\bibitem{Valiviita:2009nu}
Jussi Valiviita, Roy Maartens, and Elisabetta Majerotto.
\newblock {Observational constraints on an interacting dark energy model}.
\newblock 2009, 0907.4987.

\bibitem{Majerotto:2009np}
Elisabetta Majerotto, Jussi Valiviita, and Roy Maartens.
\newblock {Adiabatic initial conditions for perturbations in interacting dark
  energy models}.
\newblock 2009, 0907.4981.

\bibitem{Gavela:2010tm}
M.~B. Gavela, L.~Lopez~Honorez, O.~Mena, and S.~Rigolin.
\newblock {Dark Coupling and Gauge Invariance}.
\newblock {\em JCAP}, 1011:044, 2010, 1005.0295.

\bibitem{Honorez:2010rr}
Laura~Lopez Honorez, Beth~A. Reid, Olga Mena, Licia Verde, and Raul Jimenez.
\newblock {Coupled dark matter-dark energy in light of near Universe
  observations}.
\newblock {\em JCAP}, 1009:029, 2010, 1006.0877.

\bibitem{Martinelli:2010rt}
Matteo Martinelli, Laura~Lopez Honorez, Alessandro Melchiorri, and Olga Mena.
\newblock {Future CMB cosmological constraints in a dark coupled universe}.
\newblock 2010, 1004.2410.

\bibitem{LopezHonorez:2010ij}
Laura Lopez~Honorez, Olga Mena, and Grigoris Panotopoulos.
\newblock {Higher-order coupled quintessence}.
\newblock {\em Phys.Rev.}, D82:123525, 2010, 1009.5263.

\bibitem{Komatsu:2010fb}
E.~Komatsu et~al.
\newblock {Seven-Year Wilkinson Microwave Anisotropy Probe (WMAP) Observations:
  Cosmological Interpretation}.
\newblock 2010, 1001.4538.

\bibitem{Ma:1995ey}
Chung-Pei Ma and Edmund Bertschinger.
\newblock {Cosmological perturbation theory in the synchronous and conformal
  Newtonian gauges}.
\newblock {\em Astrophys. J.}, 455:7--25, 1995, astro-ph/9506072.

\bibitem{Tegmark:1996bz}
Max Tegmark, Andy Taylor, and Alan Heavens.
\newblock {Karhunen-Loeve eigenvalue problems in cosmology: How should we
  tackle large data sets?}
\newblock {\em Astrophys.J.}, 480:22, 1997, astro-ph/9603021.

\bibitem{Jungman:1995bz}
Gerard Jungman, Marc Kamionkowski, Arthur Kosowsky, and David~N. Spergel.
\newblock {Cosmological parameter determination with microwave background
  maps}.
\newblock {\em Phys.Rev.}, D54:1332--1344, 1996, astro-ph/9512139.

\bibitem{Fisher:1935bi}
Ronald~Aylmer Fisher.
\newblock {The Fiducial Argument in Statistical Inference}.
\newblock {\em Annals Eugen.}, 6:391--398, 1935.

\bibitem{Refregier:2006vt}
A.~Refregier, O.~Boulade, Y.~Mellier, B.~Milliard, R.~Pain, et~al.
\newblock {DUNE: The Dark Universe Explorer}.
\newblock 2006, astro-ph/0610062.

\bibitem{Refregier:2010ss}
A.~Refregier, A.~Amara, T.D. Kitching, A.~Rassat, R.~Scaramella, et~al.
\newblock {Euclid Imaging Consortium Science Book}.
\newblock 2010, 1001.0061.

\bibitem{dePutter:2007kf}
Roland de~Putter and Eric~V. Linder.
\newblock {To Bin or Not To Bin: Decorrelating the Cosmic Equation of State}.
\newblock {\em Astropart.Phys.}, 29:424--441, 2008, 0710.0373.

\bibitem{Heavens:2007ka}
Alan~F. Heavens, T.~D. Kitching, and L.~Verde.
\newblock {On model selection forecasting, Dark Energy and modified gravity}.
\newblock {\em Mon. Not. Roy. Astron. Soc.}, 380:1029--1035, 2007,
  astro-ph/0703191.

\bibitem{Seo:2003pu}
Hee-Jong Seo and Daniel~J. Eisenstein.
\newblock {Probing dark energy with baryonic acoustic oscillations from future
  large galaxy redshift surveys}.
\newblock {\em Astrophys.J.}, 598:720--740, 2003, astro-ph/0307460.

\bibitem{Seljak:2008xr}
Uros Seljak.
\newblock {Extracting primordial non-gaussianity without cosmic variance}.
\newblock {\em Phys.Rev.Lett.}, 102:021302, 2009, 0807.1770.

\bibitem{Seljak:2009af}
Uros Seljak, Nico Hamaus, and Vincent Desjacques.
\newblock {How to suppress the shot noise in galaxy surveys}.
\newblock {\em Phys.Rev.Lett.}, 103:091303, 2009, 0904.2963.

\bibitem{Hamaus:2011dq}
Nico Hamaus, Uros Seljak, and Vincent Desjacques.
\newblock {Optimal Constraints on Local Primordial Non-Gaussianity from the
  Two-Point Statistics of Large-Scale Structure}.
\newblock 2011, 1104.2321.

\bibitem{Huterer:2000uj}
Dragan Huterer, Lloyd Knox, and Robert~C. Nichol.
\newblock {The Angular power spectrum of EDSGC galaxies}.
\newblock {\em Astrophys.J.}, 555:547, 2001, astro-ph/0011069.

\bibitem{Tegmark:2001xb}
Max Tegmark et~al.
\newblock {The Angular power spectrum of galaxies from Early SDSS Data}.
\newblock {\em Astrophys.J.}, 571:191--205, 2002, astro-ph/0107418.

\bibitem{Hu:2003pt}
Wayne Hu and Bhuvnesh Jain.
\newblock {Joint galaxy - lensing observables and the dark energy}.
\newblock {\em Phys.Rev.}, D70:043009, 2004, astro-ph/0312395.

\bibitem{Kodama:1985bj}
Hideo Kodama and Misao Sasaki.
\newblock {Cosmological Perturbation Theory}.
\newblock {\em Prog. Theor. Phys. Suppl.}, 78:1--166, 1984.

\bibitem{Bardeen:1980kt}
James~M. Bardeen.
\newblock {Gauge Invariant Cosmological Perturbations}.
\newblock {\em Phys. Rev.}, D22:1882--1905, 1980.

\end{thebibliography}
\bibliographystyle{hunsrt}

\end{document}